\documentstyle[preprint,prd,aps,eqsecnum]{revtex}
\begin{document}
\preprint{UMN-TH-1135\hspace{-26.6mm}\raisebox{2.4ex}{SNUTP 93-14}
         \hspace{-26.6mm}\raisebox{-2.4ex}{hepth@xxx/9303131}}
\title{Schr\"{o}dinger Fields on the Plane with $[U(1)]^N$ Chern-Simons\\
Interactions and Generalized Self-dual Solitons}
\author{Chanju Kim, Choonkyu Lee}
\address{Department of Physics and Center for Theoretical Physics\\
Seoul National University, Seoul, 151-742, Korea}
\author{Pyungwon Ko, Bum-Hoon Lee\thanks{Permanent address: Department of
Physics, Hanyang University, Seoul 133-791, Korea}}
\address{School of Physics and Astronomy, University of Minnesota\\
         MN 55455, U. S. A.}
\author{and}
\author{Hyunsoo Min}
\address{Department of Physics, Seoul City University, Seoul, 130-743, Korea}
\maketitle
\newpage
\begin{abstract}
A general non-relativistic field theory on the plane with couplings to an
arbitrary number of abelian Chern-Simons gauge fields is considered.
Elementary excitations of the system are shown to exhibit fractional and
mutual statistics. We identify the self-dual systems for which certain
classical and quantal aspects of the theory can be studied in a much
simplified mathematical setting. Then, specializing to the general self-dual
system with two Chern-Simons gauge fields (and non-vanishing mutual statistics
parameter), we present a systematic analysis for the static vortexlike
classical solutions, with or without uniform background magnetic field.
Relativistic generalizations are also discussed briefly.
\end{abstract}
\pacs{11.15.Kc, 03.50.Kk}
\section{INTRODUCTION}
In two dimensional space, we can have particles obeying fractional
statistics \cite{1} and in the field theoretic context a similar effect is
generated by introducing the Chern-Simons (C-S) term \cite{2} in the action.
The C-S field theory was then found to be useful in describing the
fractional quantum Hall states \cite{3}. According to some recent suggestions
\cite{4}, a suitable generalization of it may in fact provide a unified
mathematical approach to the long-distance physics of various quantum
topological fluids. The construction involves a set of abelian gauge fields
$a^I_\mu$ ($I=1,2,\ldots,N$) with the C-S-type term
\begin{equation}   \label{1.1}
{\cal L}_g=\sum_{I,J=1}^N
  \frac{1}{2}\kappa_{IJ}\epsilon^{\mu\nu\lambda}a^I_\mu(x)a^J_\lambda(x)
\end{equation}
(here, $(\kappa_{IJ})$ is a real symmetric matrix) as the sole kinetic energy
density for them. By considering matter fields with generic gauge-invariant
couplings
to the $a^I_\mu$'s, we obtain a compound or multi-layered system which
exhibits fractional statistics for the exchange of indistinguishable particles
and mutual `statistical' interactions between particles belonging to different
species (or layers). Models of a similar nature have been considered also in
Ref. \cite{5} and, as these authors emphasize, parity need not be broken in a
field theory with an even number of C-S gauge fields. See Ref.
\cite{6} for the application to the quantum Hall effects in the double-layer
electron system.

The purpose of this paper is two-fold. Firstly, we clarify the precise nature
of a non-relativistic quantum field theory incorporating the above form of
C-S-type interaction. In particular we find explicitly the
corresponding first quantized description in the general $n$-body sector. In
the latter description, fractional and mutual statistics for the particles are
the manifestations of the Aharonov-Bohm effect involving a combination of
fictitious charges and localized fluxes affixed to them. This is described in
Sec. II. (For a complementary discussion on mutual statistics from the braid
group viewpoint, readers are referred to Ref. \cite{7}).

Secondly, in Sec. III, we identify the corresponding {\it self-dual} system
(with an arbitrary background magnetic field), which has a simpler
mathematical structure than the generic case due to hidden supersymmetry
\cite{8}. For instance, thanks to the supersymmetry, one can construct the
exact many-body ground state wave function in this case. (See Ref. \cite{9}
for related discussions). In this paper, however, we concentrate on the
analysis of static soliton solutions to the classical field equations that
follow, i.e., look for a generalization of the Jackiw-Pi solutions
\cite{10,11}.
More general types of vortex solitons, which are likely to exhibit
fractional and mutual statistics themselves, are found. The self-duality
equations for our model share certain common elements with those for self-dual
non-abelian C-S vortices discussed recently \cite{12,13}. For
instance, the Toda-type equation
\begin{equation}  \label{1.2}
\nabla^2\ln|\phi_p({\bf r})|^2=-K_{pp'}|\phi_{p'}({\bf r})|^2
\end{equation}
has a prominent role in both cases. But note that, in our model, $K=(K_{pp'})$
need {\it not\/} be equal to the Cartan matrix of a certain Lie algebra and
very little is known for this case. Concentrating on the
case of two C-S gauge fields, we will present a fairly
detailed study on the nature of possible self-dual vortex solutions, with or
without uniform background magnetic field. For the case with non-zero magnetic
field the self-duality equations were also touched upon in Ref. \cite{6}, but
there is only a minimal overlap between the latter work and ours.

Fully relativistic self-dual systems, including several abelian C-S
gauge fields, are also possible and we briefly discuss them in Sec. IV. Nature
of static soliton solutions is discussed in the special case of these models.
Section V contains a summary and discussion of our work. There are two
appendices. In the first we provide the derivations of certain formulas
appearing in Sec. II. The second appendix contains the index theorem analysis
for the self-dual system being treated in this paper.

\section{NON-RELATIVISTIC QUANTUM FIELD THEORY}
Choosing the basis where the matrix $(\kappa_{IJ})$ (see Eq. (\ref{1.1})) is
diagonal, let
us consider the non-relativistic C-S gauge field theory defined by
the Lagrangian density\footnote{Note that $\mu,\nu,\lambda=0,1$ or 2 and our
(2+1)-D metric is given as $\eta_{\mu\nu}=\mbox{diag.}(-1,1,1)$.}
\begin{eqnarray}
{\cal L}&=&\sum_{I=1}^N\frac{1}{2}\kappa_I\epsilon^{\mu\nu\lambda}
     a^I_\mu\partial_\nu a^I_\lambda
    +\sum_{p=1}^M\left\{i\hbar\Psi_p^\dagger
    \left(\frac{\partial}{\partial t}-\frac{i}{\hbar}
     \sum_{I=1}^Nq_p^Ia^I_0\right)\Psi_p\right.\nonumber\\
    & &\left.-\frac{\hbar^2}{2m_p}\left|
    \left(\nabla-\frac{i}{\hbar c}\sum_{I=1}^Nq_p^I{\bf a}^I
    -\frac{i}{\hbar c}e_p{\bf A}^{\rm ex}\right)\Psi_p\right|^2\right\}
    -U(\Psi^\dagger,\Psi),
\end{eqnarray}
where $\Psi_p$ ($p=1,\ldots,M$) denote $M$ different bosonic (fermionic) fields
satisfying the equal-time (anti-)commutation relations
\begin{eqnarray}  \label{2.2}
&&[\Psi_p({\bf r},t),\Psi_{p'}({\bf r}',t)]_{\mp}
=[\Psi_p^\dagger({\bf r},t),\Psi_{p'}^\dagger({\bf r}',t)]_{\mp}=0,\nonumber\\
&&[\Psi_p({\bf r},t),\Psi_{p'}^\dagger({\bf r}',t)]_{\mp}
 =\delta_{pp'}\delta^2({\bf r}-{\bf r}').
\end{eqnarray}
[The suffix $-(+)$ refers to the commutator (anticommutator)]. This system
possesses $[U(1)]^N$ local gauge invariance in connection with $N$ independent
C-S gauge fields $a^I_\mu$, and we have included the external electromagnetic
field ${\bf A}^{\rm ex}$ for the sake of generality. Excluding the
pathological cases, we will below assume that all $\kappa_I$'s are non-zero,
and $M\ge N$, i.e., the number of the C-S fields does not exceed that of
the matter fields. Also, for definiteness, we shall take the
potential $U(\Psi^\dagger,\Psi)$ to have the general form
\begin{eqnarray}
U&=&\sum_p V_p({\bf r},t)\Psi_p^\dagger({\bf r},t)\Psi_p({\bf r},t) \nonumber\\
 &&+\frac{1}{2}\sum_{p,p'}\int\! d^2\!{\bf r}'\,
    \Psi_p^\dagger({\bf r},t)\Psi_{p'}^\dagger({\bf r}',t)
    V_{pp'}({\bf r}-{\bf r}')\Psi_{p'}({\bf r}',t)\Psi_p({\bf r},t),
\end{eqnarray}
where $V_{pp'}({\bf r}-{\bf r}')=V_{p'p}({\bf r}'-{\bf r})$.

The stationary action principle for varying $a^I_0$ yields the Gauss laws
($i,j=1$ or 2)
\begin{equation}  \label{2.4}
b^I\equiv\epsilon^{ij}\nabla_ia^I_j
   =-\frac{1}{\kappa_I}\sum_pq_p^I\Psi_p^\dagger\Psi_p\,.
\end{equation}
These, together with the Coulomb gauge conditions $\nabla_ia^I_i=0$, then
determine ${\bf a}^I({\bf r},t)$ in terms of the matter densities
$\rho_p({\bf r},t)\equiv \Psi_p^\dagger({\bf r},t)\Psi_p({\bf r},t)$,
\begin{equation}  \label{2.5}
a^I_i({\bf r},t)=\epsilon^{ij}\nabla_j\frac{1}{\kappa_I}
  \sum_pq_p^I\int\! d^2\!{\bf r}'\, G({\bf r}-{\bf r}')\rho_p({\bf r}',t),
\end{equation}
where $G$ is the Green's function for the 2-D Laplacian, i.e.,
\begin{eqnarray}  \label{2.6}
\nabla^2 G({\bf r}-{\bf r}')&=&\delta^2({\bf r}-{\bf r}'),\nonumber\\
G({\bf r}-{\bf r}')&=&\frac{1}{2\pi}\ln{|{\bf r}-{\bf r}'|}.
\end{eqnarray}
Now the Hamiltonian operator of the system can be identified with
\begin{equation} \label{2.7}
H=\int\! d^2\!{\bf r}\,\left\{\sum_{p=1}^M\frac{\hbar^2}{2m_p}
  \mbox{\boldmath $\Pi$}_p^\dagger({\bf r},t)
  \cdot\mbox{\boldmath $\Pi$}_p({\bf r},t)+U(\Psi^\dagger,\Psi)\right\},
\end{equation}
where
\begin{eqnarray}   \label{2.8}
\mbox{\boldmath $\Pi$}_p({\bf r},t)&=&\left[\nabla-\frac{i}{\hbar c}
     \sum_Iq_p^I{\bf a}^I({\bf r},t)-\frac{i}{\hbar c}e_p{\bf A}^{\rm ex}
     ({\bf r},t)\right]\Psi_p({\bf r},t)\nonumber\\
    &\equiv&{\bf D}\Psi_p({\bf r},t)
\end{eqnarray}
with the fields ${\bf a}^I$ expressed in terms of
$\Psi^\dagger$ and $\Psi$ through Eq. (\ref{2.5})

What we have in Eq.(\ref{2.7}) is the {\it properly ordered\/} Hamiltonian,
and the corresponding operator field equations are
\begin{eqnarray}  \label{2.9}
i\hbar\frac{\partial\Psi_p({\bf r},t)}{\partial t}
 &=&[\Psi_p({\bf r},t),H]\nonumber\\
 &\equiv&\sum_Iq_p^Ia^I_0({\bf r},t)\Psi_p({\bf r},t)-\frac{\hbar^2}{2m_p}
  {\bf D}^2\Psi_p({\bf r},t)+V_p({\bf r},t)\Psi_p({\bf r},t)\nonumber\\
 & &+\int\! d^2\!{\bf r}'\,\Psi_{p'}^\dagger({\bf r}',t)
    V_{pp'}({\bf r}-{\bf r}')\Psi_{p'}({\bf r}',t)\Psi_p({\bf r},t)+{\cal R},
\end{eqnarray}
where $\cal R$, the quantum correction from operator ordering as first
discussed in Ref. \cite{11}, is specified as
\begin{equation}
{\cal R}=\sum_{p'}\frac{1}{2m_{p'}c^2}
        \left(\sum_I\frac{q_{p'}^Iq_p^I}{\kappa_I}\right)^2
   \int\! d^2\!{\bf r}'\,\left(\frac{1}{4\pi^2}
  \frac{1}{|{\bf r}-{\bf r}'|^2}\right)\rho_{p'}({\bf r}',t)\Psi_p({\bf r},t),
\end{equation}
and the operators $a^I_0({\bf r},t)$ represent the solutions to the equations
\begin{equation}
\kappa_I\epsilon^{ij}\nabla_j a^I_0({\bf r},t)+i\hbar\sum_p\frac{q_p^I}{2m_pc}
  \left[(D_i\Psi_p)^\dagger({\bf r},t)\Psi_p({\bf r},t)
  -\Psi_p^\dagger({\bf r},t)(D_i\Psi_p)({\bf r},t)\right]=0\,,
\end{equation}
or, more explicitly,
\begin{eqnarray} \label{2.12}
a^I_0({\bf r},t)&=&\sum_p\frac{q_p^I}{\kappa_Ic}\epsilon^{ij}\nabla_j
   \int\! d^2\!{\bf r}'\, G({\bf r}-{\bf r}')J_{pi}({\bf r}',t)\,,\nonumber\\
J_{pi}({\bf r},t)&\equiv&\frac{i\hbar}{2m_p}
    \left[(D_i\Psi_p)^\dagger({\bf r},t)\Psi_p({\bf r},t)
   -\Psi_p^\dagger({\bf r},t)(D_i\Psi_p)({\bf r},t)\right]\,.
\end{eqnarray}
{}From the (anti-)commutation relations (\ref{2.2}) it follows that
\begin{equation}  \label{2.13}
[\Psi_p({\bf r}',t),a^I_i({\bf r},t)]=\epsilon^{ij}\nabla_j
   \frac{1}{\kappa_I}q_p^I G({\bf r}-{\bf r}')\Psi_p({\bf r}',t),
\end{equation}
and, in evaluating the commutator needed to derive Eq. (\ref{2.9}), we have
taken (following Ref. \cite{11}) that
\begin{equation}
[\Psi_{p'}({\bf r}',t), a^I_i({\bf r},t)]|_{{\bf r}'={\bf r}}=0\,,
\end{equation}
i.e., the quantity $\epsilon^{ij}\nabla_jG({\bf r}-{\bf r}')$ at the
coincidence limit
${\bf r}'={\bf r}$ has been prescribed to be zero. Now we denote
\begin{equation}  \label{2.15}
\sum_{I}\frac{q_p^Iq_{p'}^I}{\kappa_I}\equiv\beta_{pp'}(=\beta_{p'p})
\end{equation}
and then
\begin{eqnarray}
D_i\Psi_p({\bf r},t)&=&\left\{\nabla_i-\frac{i}{\hbar c}\epsilon^{ij}\nabla_j
  \sum_{p'}\beta_{pp'}\int\! d^2\!{\bf r}'\,
  G({\bf r}-{\bf r}')\rho_{p'}({\bf r}',t)\right.\nonumber\\
& &-\frac{i}{\hbar c}e_p{\bf A}^{\rm ex}({\bf r},t)\Biggr\}\Psi_p({\bf r},t),\\
\sum_{I}q_p^Ia^I_0({\bf r},t)
  &=&\sum_{p'}\frac{\beta_{pp'}}{c}\epsilon^{ij}\nabla_j
  \int\! d^2\!{\bf r}'\, G({\bf r}-{\bf r}')J_{p'i}({\bf r}',t),
\end{eqnarray}
showing that the parameters $q_p^I$'s, $\kappa_I$'s enter the field
equations only
through the $\beta_{pp'}$'s. From this we infer that theories with different
$q_p^I$'s and $\kappa_I$'s but with the same values for the $\beta_{pp'}$'s are
physically equivalent. A simple physical interpretation for the $\beta_{pp'}$'s
will be given later.

To see clearly the physical content of the above non-relativistic quantum
field theory, we will now derive the equivalent first-quantized description.
For the two-particle sector of a one-layer system (i.e., a single matter
field), this has been explicitly performed in Ref. \cite{11}. The general
$n$-body sector with the Hamiltonian (\ref{2.7}) will be considered
here\footnote{After the completion of this paper we learned that C.-L. Ho and
Y. Hosotani \cite{ho} previously considered the Schr\"{o}dinger equation for
$n$ anyons (on a torus), starting from the corresponding C-S field theory. Ours
is more explicit and also deals with more general Hamiltonian, and so we
include this discussion for the sake of completeness.}. Let $|\Phi\rangle$
denote any Heisenberg-picture state vector with the total number of particles
equal to $n=\sum_{p=1}^Mn_p$. Then the corresponding many-particle
Schr\"odinger wave function is given by
\begin{equation}   \label{2.18}
\Phi({\bf r}_1^{(1)},\ldots,{\bf r}_{n_1}^{(1)},\ldots,{\bf r}_1^{(M)},\ldots,
                                                    {\bf r}_{n_M}^{(M)},t)
  =\left\langle 0\left|\prod_{p=1}^M\frac{1}{\sqrt{n_p !}}\Psi_p
   ({\bf r}_1^{(p)},t)
   \cdots\Psi_p({\bf r}_{n_p}^{(p)},t)\right|\Phi\right\rangle
\end{equation}
where the vacuum $|0\rangle$ satisfies the conditions
$\Psi_p({\bf r},t)|0\rangle=0$ for any
$p=1,\ldots,M$. Now, using the field equations (\ref{2.9}), we have
\begin{eqnarray}  \label{2.19}
&&i\hbar\frac{\partial}{\partial t}
\Phi({\bf r}_1^{(1)},\ldots,{\bf r}_{n_1}^{(1)},\ldots,{\bf r}_1^{(M)},
                                   \ldots,{\bf r}_{n_M}^{(M)},t)=\sum_p
  \left\langle 0\left|\left\{\frac{1}{\sqrt{n_1 !}}\Psi_1({\bf r}_1^{(1)},t)
           \cdots\Psi_1({\bf r}_{n_1}^{(1)},t)\right\}\right.\right.\nonumber\\
& &\hspace{14mm}\cdots\left\{\sum_{k=1}^{n_p}\frac{1}{\sqrt{n_p!}}
   \Psi_p({\bf r}_1^{(p)},t)\cdots
   \left(i\hbar\frac{\partial}{\partial t}\Psi_p({\bf r}_k^{(p)},t)
   \right)\cdots\Psi_p({\bf r}_{n_p}^{(p)},t)\right\}\nonumber\\
& &\hspace{14mm}\cdots\left.\left.\left\{\frac{1}{\sqrt{n_M!}}
    \Psi_M({\bf r}_1^{(M)},t)\cdots\Psi_M({\bf r}_{n_M}^{(M)},t)\right\}
  \right|\Phi\right\rangle\nonumber\\
&&\hspace{5mm}=A+B+C+D
\end{eqnarray}
with
\begin{eqnarray}
A&=&\sum_p\left\langle 0\left|\cdots\left\{
 \sum_{k=1}^{n_p}\frac{1}{\sqrt{n_p!}}\Psi_p({\bf r}_1^{(p)},t)
                                        \right.\right.\right.\nonumber\\
 & &\hspace{10mm}\left.\cdots
   \left(\sum_{p'}\frac{\beta_{pp'}}{c}\epsilon^{ij}\nabla_j^{(p,k)}\!\!
   \int\! d^2\!{\bf r}'\, G({\bf r}_k^{(p)}-{\bf r}')
   J_{p'i}({\bf r}',t)\right)\Psi_p({\bf r}_k^{(p)},t)
   \cdots\Psi_p({\bf r}_{n_p}^{(p)},t)\right\}\nonumber\\
 & &\hspace{10mm}\cdots\Biggl|\Phi\Biggr\rangle\,,\label{2.20}\\
B&=&\sum_p\left\langle 0\left|\cdots\left\{
 \sum_{k=1}^{n_p}\frac{1}{\sqrt{n_p!}}\Psi_p({\bf r}_1^{(p)},t)
 \right.\right.\right.\nonumber  \\
 & &\hspace{10mm}\cdots
   \left(-\frac{\hbar^2}{2m_p}\left[\nabla_i^{(p,k)}
   -\frac{i}{\hbar c}\epsilon^{ij}\nabla_j^{(p,k)}
    \sum_{p'}\beta_{pp'}\int\! d^2\!{\bf r}'\,
   G({\bf r}_k^{(p)}-{\bf r}')\rho_{p'}({\bf r}',t)\right.\right. \nonumber\\
 &&\left.\left.\left.\left.\hspace{8mm}
  -\frac{i}{\hbar c}e_pA^{\rm ex}_i({\bf r}_k^{(p)},t)\Biggr]^2
   \Psi_p({\bf r}_k^{(p)},t)\right)\cdots\Psi_p({\bf r}_{n_p}^{(p)},t)
   \right\}\cdots\right|\Phi\right\rangle\,,\label{2.21}\\
C&=&\sum_p\left\langle 0\left|\cdots\left\{
 \sum_{k=1}^{n_p}\frac{1}{\sqrt{n_p!}}\Psi_p({\bf r}_1^{(p)},t)\cdots
 \Biggl(V_p({\bf r}_k^{(p)},t)\Psi_p({\bf r}_k^{(p)},t)
  \right.\right.\right. \nonumber\\
 &&\left.\left.\left.\left.
   +\sum_{p'}\int\! d^2\!{\bf r}'\,\Psi_{p'}^\dagger({\bf r}',t) V_{pp'}
   ({\bf r}_k^{(p)}-{\bf r}')\Psi_{p'}({\bf r}',t)\Psi_p
   ({\bf r}_k^{(p)},t)\right)\cdots\Psi_p({\bf r}_{n_p}^{(p)},t)\right\}
  \cdots\right|\Phi\right\rangle\,,\nonumber\\
  &&\label{2.22}\\
D&=&\sum_p\left\langle 0\left|\cdots\left\{
 \sum_{k=1}^{n_p}\frac{1}{\sqrt{n_p!}}\Psi_p({\bf r}_1^{(p)},t)
                                             \right.\right.\right.\nonumber\\
 & &\left.\hspace{10mm}\cdots
   \left(\sum_{p'}\frac{1}{2m_{p'}c^2}\beta_{pp'}^2
   \int\! d^2\!{\bf r}'\,\frac{1}{4\pi^2}
   \frac{1}{|{\bf r}_k^{(p)}-{\bf r}'|^2}\rho_{p'}({\bf r}',t)
   \Psi_p({\bf r}_k^{(p)},t)\right)
   \cdots\Psi_p({\bf r}_{n_p}^{(p)},t)\right\}\nonumber\\
 &&\hspace{9mm}\cdots\Biggl|\Phi\Biggr\rangle\,.
\end{eqnarray}
where $\nabla_i^{(p,k)}$ denotes the derivative with respect to
${\bf r}_k^{(p)}$.

The contributions designated as $C$ and $D$ above have the structure of the
standard one-body and two-body interaction terms. So we may immediately rewrite
them as \cite{fetter}
\begin{eqnarray} \label{2.24}
C&=&\left\{\sum_{(p,k)}V_p({\bf r}_k^{(p)},t)+\frac{1}{2}
\sum_{(p,k)}\sum_{(p',k')\neq(p,k)}V_{pp'}({\bf r}_k^{(p)}-{\bf r}_{k'}^{(p')})
  \right\}\Phi({\bf r}_1^{(1)},\ldots,{\bf r}_{n_M}^{(M)},t)\,,\ \\
D&=&\left\{\frac{1}{2}\sum_{(p,k)}\sum_{(p',k')\neq(p,k)}\frac{1}{2m_{p'}c^2}
    \beta_{pp'}^2\frac{1}{4\pi^2}\frac{1}{|{\bf r}_k^{(p)}
    -{\bf r}_{k'}^{(p')}|^2}\right\}
    \Phi({\bf r}_1^{(1)},\ldots,{\bf r}_{n_M}^{(M)},t)\,,   \label{2.25}
\end{eqnarray}
where the sum $\frac{1}{2}\sum_{(p,k)}\sum_{(p',k')\neq(p,k)}$ has the effect
of taking in once every different set of pairs $(p,k)$ and $(p',k')$,
excluding the case with $p'=p$ {\it and\/} $k'=k$. On the other hand, we show
in Appendix A that the contributions B and A above can in fact be expressed as
\begin{eqnarray}
B&=&\sum_{(p,k)}\left(-\frac{\hbar^2}{2m_p}\right)\left[\nabla_i^{(p,k)}
 -\frac{i}{\hbar c}\epsilon^{ij}\nabla_j^{(p,k)}\sum_{(p',k')<(p,k)}\beta_{pp'}
G({\bf r}_k^{(p)}-{\bf r}_{k'}^{(p')})
       -\frac{i}{\hbar c}e_pA^{\rm ex}_i({\bf r}_k^{(p)},t)\right]^2\nonumber\\
  &&\hspace{26mm}\cdot\Phi({\bf r}_1^{(1)},\ldots,
    {\bf r}_{n_M}^{(M)},t)\,,\label{2.27}\\
A&=&\sum_{(p,k)}\sum_{(p',k')<(p,k)}\left(-\frac{i\hbar}{m_{p'}c}\right)
 \beta_{pp'}\epsilon^{ij}\nabla_j^{(p,k)}G({\bf r}_k^{(p)}-{\bf r}_{k'}^{(p')})
    \Biggl[\nabla_i^{(p',k')} \nonumber\\
 &&-\frac{i}{\hbar c}\epsilon^{il}\nabla_l^{(p',k')}
   \sum_{(p'',k'')<(p,k)\atop (p'',k'')\neq(p',k')}\beta_{p'p''}
   G({\bf r}_{k'}^{(p')}-{\bf r}_{k''}^{(p'')})
   -\frac{i}{\hbar c}e_{p'}A^{\rm ex}_i({\bf r}_{k'}^{(p')},t)\Biggr]
   \Phi({\bf r}_1^{(1)},\ldots,{\bf r}_{n_M}^{(M)},t),\nonumber\\
 &&\label{2.30}
\end{eqnarray}
where $\sum_{(p',k')<(p,k)}$ denotes the sum over all indices $(p',k')$ that
appear on the left of ${\bf r}_k^{(p)}$ in the arrangement
$({\bf r}_1^{(1)},\ldots,{\bf r}_{n_1}^{(1)},\ldots,{\bf r}_1^{(p)},
\ldots,{\bf r}_k^{(p)},
\ldots,{\bf r}_{n_p}^{(p)},\ldots,{\bf r}_1^{(M)},\ldots,{\bf r}_{n_M}^{(M)})$.
In Appendix A it is further shown (after a bit involved manipulations) that
the contributions $A$, $B$ and $D$ above combine to give a surprisingly simple
expression, namely,
\begin{eqnarray}  \label{2.31}
A+B+D&=&\sum_{(p,k)}\left(-\frac{\hbar^2}{2m_p}\right)
    \left[\nabla_i^{(p,k)}-\frac{i}{\hbar c}
    \epsilon^{ij}\nabla_j^{(p,k)}\left(\sum_{(p',k')\neq(p,k)}\beta_{pp'}
    G({\bf r}_k^{(p)}-{\bf r}_{k'}^{(p')})\right)\right.\nonumber\\
 &&\hspace{26mm}-\frac{i}{\hbar c}e_pA^{\rm ex}_i({\bf r}_k^{(p)},t)\Biggr]^2
     \Phi({\bf r}_1^{(1)},\ldots,{\bf r}_{n_M}^{(M)},t).
\end{eqnarray}
Using this result in Eq. (\ref{2.19}), we then find that the appropriate
many-particle Schr\"{o}dinger equation reads
\begin{eqnarray}  \label{2.35}
i\hbar\frac{\partial}{\partial t}&&\Phi({\bf r}_1^{(1)},
  \ldots,{\bf r}_{n_M}^{(M)},t)=
  \left\{\sum_{(p,k)}\left(-\frac{\hbar^2}{2m_p}\right)\left[\nabla^{(p,k)}
  -\frac{i}{\hbar c}{\bf\cal A}_{(p,k)}({\bf r}_1^{(1)},
  \ldots,{\bf r}_{n_M}^{(M)})-\frac{i}{\hbar c}e_p{\bf A}^{\rm ex}
  ({\bf r}_k^{(p)},t)\right]^2\right.\nonumber\\
&&\left.+\sum_{(p,k)}V_p({\bf r}_k^{(p)},t)
  +\frac{1}{2}\sum_{(p,k)}\sum_{(p',k')\neq(p,k)}
       V_{pp'}({\bf r}_k^{(p)}-{\bf r}_{k'}^{(p')})\right\}
                        \Phi({\bf r}_1^{(1)},\ldots,{\bf r}_{n_M}^{(M)},t),
\end{eqnarray}
where we have defined
\begin{eqnarray}  \label{2.36}
{\cal A}_{(p,k)i}({\bf r}_1^{(1)},\ldots,{\bf r}_{n_M}^{(M)})
&=&\epsilon^{ij}\nabla_j^{(p,k)}\left(\sum_{(p',k')\neq(p,k)}\beta_{pp'}
   G({\bf r}_k^{(p)}-{\bf r}_{k'}^{(p')})\right)\nonumber\\
&=&\epsilon^{ij}\nabla_j^{(p,k)}\left(\frac{1}{2}\sum_{(p',k')}
   \sum_{(p'',k'')\neq(p',k')}
   \beta_{p'p''}G({\bf r}_{k'}^{(p')}-{\bf r}_{k''}^{(p'')})\right).\
\end{eqnarray}
The wave function $\Phi$ should be single-valued with respect to every particle
coordinate and satisfy the symmetry requirement appropriate to bosons or
fermions,
\begin{equation}  \label{2.37}
\Phi(\ldots,{\bf r}_{k_1}^{(p)},\ldots,{\bf r}_{k_2}^{(p)},\ldots)
   =\pm\Phi(\ldots,{\bf r}_{k_2}^{(p)},\ldots,{\bf r}_{k_1}^{(p)},\ldots).
\end{equation}

The Schr\"odinger equation (\ref{2.35}) provides an equivalent
first-quantized description for the quantum field theory defined by the
ordered Hamiltonian operator (\ref{2.7}). The entire effect of the
C-S interactions now enters through the vector potentials
${\cal A}_{(p,k)}$, which can be related to the induced Aharonov-Bohm-type
interactions between the charge-flux composites of suitable nature. In
particular, the form (\ref{2.36}) implies that the Aharonov-Bohm interaction
strength between the two particles carrying the labels $p$ and $p'$ is equal
to $-\beta_{pp'}$. A straightforward interpretation of this is as follows. In
view of the fact that we had $N$ abelian C-S gauge fields $a^I_\mu$
($I=1,\ldots,N$), we associate with a type-$p$ particle an $N$-tuple of
charges $(q_p^1,\ldots,q_p^N)$ and also an $N$-tuple of corresponding fluxes
$(-\frac{q^1_p}{\kappa_1},\ldots,-\frac{q^N_p}{\kappa_N})$ in accordance
with the Gauss laws (\ref{2.4}). Then, given two particles from type-$p$ and
type-$p'$ each, the expected strength of the Aharonov-Bohm interaction will be
$\sum_Iq_p^I(-\frac{q_{p'}^I}{\kappa_I})=-\beta_{pp'}$, in agreement with
the observation we just made. [Actually, with the Aharonov-Casher
interaction \cite{14} taken as an {\it additional\/} effect to the
Aharonov-Bohm interaction, the flux affixed to a $p$-type particle will have to
read $(-\frac{q^1_p}{2\kappa_1},\ldots,-\frac{q^N_p}{2\kappa_N})$. Note
that, in interpreting the many-particle Schr\"{o}dinger equation (\ref{2.35}),
we are in no way bound by the Gauss laws (\ref{2.4}).]

We here remark that the above interpretation, based on {\it $N$ distinct
$U(1)$ charges and corresponding fluxes,} is not the only possible. We will
illustrate this phenomenon through a closer look at the $N=1$ and $N=2$ cases.
With $N=1$ (i.e. one C-S gauge field only) but arbitrary number of matter
fields, we may write $\beta_{pp'}=\frac{q_pq_p'}{\kappa}$ and so assign flux
$-\frac{q_p}{\kappa}$ to a particle of charge $q_p$; here, the charge-flux
ratio is necessarily the same for all particle species. With $N=2$
(i.e. two C-S gauge fields), on the other hand, we have the formula
\begin{equation}  \label{2.38}
\beta_{pp'}=\frac{q^1_pq_{p'}^1}{\kappa_1}+\frac{q^2_pq_{p'}^2}{\kappa_2},
  \hspace{5mm}(p,p'=1,2,\ldots,M)
\end{equation}
from which we immediately derive the results
\begin{eqnarray} \label{2.39}
\beta_{pp'}^2&\le&\beta_{pp}\beta_{p'p'},\hspace{7mm}
    \mbox{if $\kappa_1\kappa_2>0$}\nonumber\\
\beta_{pp'}^2&\ge&\beta_{pp}\beta_{p'p'},\hspace{7mm}
    \mbox{if $\kappa_1\kappa_2<0$}
\end{eqnarray}
for any given $p$, $p'$. Equation (\ref{2.39}) puts a restriction on the
possible values of the $\beta_{pp'}$ for $M>2$ ( to be realizable by a
$U(1)\times U(1)$ C-S field theory), and here the sign of $\kappa_1\kappa_2$
matters also. As we described in the previous paragraph, this system may be
related to that of composites carrying appropriate 2-vector charges and
corresponding 2-vector fluxes. But, for the case with $\kappa_1\kappa_2<1$, an
alternative, in some sense simpler, interpretation is also available.
Specifically, we assign (scalar) charge
$\widetilde q_p=\frac{q^1_p}{\sqrt{|\kappa_1|}}
  -\frac{q_p^2}{\sqrt{|\kappa_2|}}$ and flux
$\widetilde{\phi}_p=\pm(\frac{q^1_p}{2\sqrt{|\kappa_1|}}
   +\frac{q_p^2}{2\sqrt{|\kappa_2|}})$
to a type $p$ particle, and then the
Aharonov-Bohm and Aharonov-Casher interactions between two particles from
type-$p$ and type-$p'$ will have the net strength
\begin{equation}
\widetilde{q}_p\widetilde\phi_{p'}+\widetilde q_{p'}\widetilde\phi_p
  =\pm\left[\frac{q^1_pq_{p'}^1}{|\kappa_1|}
   -\frac{q_p^2q_{p'}^2}{|\kappa_2|}\right]\,,
\end{equation}
i.e., equal to $\beta_{pp'}$ (see Eq. (\ref{2.38})) under the restriction
$\kappa_1\kappa_2<0$. This shows that a multi-component system of charge-flux
composites, with {\it different\/} charge-flux ratios for individual
components, can equivalently be represented by a $U(1)\times U(1)$
C-S field theory. We have an exceptional situation if
\begin{equation}  \label{2.41}
\beta_{pp'}^2-\beta_{pp}\beta_{p'p'}=0\hspace{6mm}\mbox{(for every $p$, $p'$)},
\end{equation}
and for this case the charge-flux ratios $\widetilde q_p/\widetilde\phi_p$
become independent of
$p$. When the $\beta_{pp'}$'s satisfy the  conditions (\ref{2.41}), the
equivalent $U(1)\times U(1)$ C-S theory  constructed according to the
above correspondence is effectively reduced to a $U(1)$ C-S theory;
i.e., from the two C-S gauge fields, it is only their particular
linear combination that has a dynamical role. [Incidentally, if
$\kappa_1\kappa_2<0$, we will always be allowed to set
$\kappa_1=-\kappa_2=\kappa$ thanks to the rescaling
freedom of the C-S fields. Then we may introduce new C-S
fields $v^{(1)}_\mu$, $v^{(2)}_\mu$ by
\begin{equation}   \label{2.42}
a_{1\mu}=\frac{1}{\sqrt{2}}(v^{(1)}_\mu+v^{(2)}_\mu),\hspace{5mm}
a_{2\mu}=\frac{1}{\sqrt{2}}(v^{(1)}_\mu-v^{(2)}_\mu),
\end{equation}
and in terms of these fields the C-S Lagrangian becomes
\begin{equation}
{\cal L}_g
 =\kappa\epsilon^{\mu\nu\lambda}v^{(1)}_\mu\partial_\nu v^{(2)}_\lambda\,.
\end{equation}
The Lagrangian of this form has been considered recently by Wilczek \cite{5}.]
It should be an interesting mathematical exercise to extend the above
discussion to
the case of general $N$, but it is not pursued in this paper.

As is well-known, the Aharonov-Bohm interaction affects the statistical
character of the particles involved. To see this, observe that the vector
potential in Eq. (\ref{2.36}) is locally a pure gauge,
\begin{eqnarray}
&&{\cal A}_{(p,k)i}({\bf r}_1^{(1)},\ldots,{\bf r}_{n_M}^{(M)})
  =-\nabla_i^{(p,k)}\Lambda\,,\nonumber\\
&&\Lambda=\frac{1}{2}\sum_{(p',k')}\sum_{(p'',k'')\neq(p',k')}\beta_{p'p''}
 \frac{1}{2\pi}\tan^{-1}\left(
 \frac{y_{k'}^{(p')}-y_{k''}^{(p'')}}{x_{k'}^{(p')}-x_{k''}^{(p'')}}\right)\,.
\end{eqnarray}
So, by redefining the single-valued wave function $\Phi$ according to
\begin{equation}  \label{2.45}
\Phi({\bf r}_1^{(1)},\ldots,{\bf r}_{n_M}^{(M)},t)
  =e^{-\frac{i}{\hbar c}\Lambda}\bar\Phi({\bf r}_1^{(1)},\ldots,
   {\bf r}_{n_M}^{(M)},t)\,,
\end{equation}
we may have the gauge potentials ${\cal A}_{(p,k)}$ disappear in the
Schr\"{o}dinger equation for $\bar\Phi$. But the gauge function $\Lambda$ here
being multi-valued in general, $\bar\Phi$ will have to be multi-valued (for a
single-valued $\Phi$) and so satisfy highly non-trivial boundary conditions
that
are sensitive to the $\beta_{p'p''}$'s. To be explicit, consider exchanging the
positions of two identical particles, say ${\bf r}_{k_1}^{(p)}$ and
${\bf r}_{k_2}^{(p)}$, with the function
$\bar\Phi(\ldots,{\bf r}_{k_1}^{(p)},\ldots,{\bf r}_{k_2}^{(p)},\ldots)$ along
a certain closed path $C$ (see Fig. 1). Then from Eqs. (\ref{2.37}) and
(\ref{2.45}), the resulting expression should differ from the original by the
phase factor
$\pm e^{\frac{i}{2\hbar c}\beta_{pp}}e^{\frac{i}{\hbar c}\gamma(C)}$, where
$\gamma(C)$ is equal to the sum of the $\beta_{pp'}$'s over the index set
$(p',k')$ associated with the ${\bf r}_{k'}^{(p')}$'s in the interior of $C$.
[Here an implicit assumption is that no position variable other than
${\bf r}_{k_1}^{(p)}$ and ${\bf r}_{k_2}^{(p)}$ takes values on $C$.]
Also note that as we allow a specific position variable ${\bf r}_k^{(p)}$
to be taken along a closed path
$C$, the initial and final expressions of $\bar\Phi$ should differ by a phase
$e^{\frac{i}{\hbar c}\gamma(C)}$, with $\gamma(C)$ defined as above.
If all $\beta_{pp'}$'s with $p\neq p'$ vanish, the system is that
of $M$ species of fractional-statistics-obeying particles. In the terminology
of Wilczek \cite{5}, the $\beta_{pp'}$'s with $p\neq p'$ are relevant for {\it
mutual\/} statistics, while the $\beta_{pp}$'s are responsible for more usual
fractional statistics. We close this section with the remark that, because of
the highly non-trivial nature of the boundary conditions satisfied by
$\bar\Phi$, the regular gauge description based on a single-valued function
$\Phi$ should be preferred for all more explicit studies.

\section{SELF-DUAL SYSTEMS AND SOLITON SOLUTIONS}
The system described by the field theory of Sec. II is highly nontrivial, and
it is extremely difficult to obtain any concrete information on its behavior.
Naturally, one might then ask whether there exists a certain special choice of
the potential $U(\Psi^\dagger,\Psi)$ for which the mathematical treatment of
the system becomes more tractable. This leads us to consider the so-called
self-dual system, which is based on the potential
\begin{eqnarray} \label{3.1}
U&=&\sum_p \left(-\frac{\hbar}{2m_pc}\sigma_pe_pB^{\rm ex}({\bf r})\right)
   \Psi_p^\dagger({\bf r},t)\Psi_p({\bf r},t)\nonumber\\
 & &+\frac{1}{2}\sum_{p,p'}\left(\frac{\hbar}{2m_pc}\sigma_p
  +\frac{\hbar}{2m_{p'}c}\sigma_{p'}\right)
  \beta_{pp'}\Psi_p^\dagger({\bf r},t)\Psi_{p'}^\dagger
  ({\bf r},t)\Psi_{p'}({\bf r},t)\Psi_p({\bf r},t)\nonumber\\
 &&
\end{eqnarray}
where $B^{\rm ex}({\bf r})\equiv\epsilon^{ij}\nabla_iA^{\rm ex}_j({\bf r})$
and $\sigma_p=1$ or $-1$
(for each $p$, independently). For this choice of the potential, the
many-particle Schr\"{o}dinger equation (\ref{2.35}) can be cast into the form
\begin{eqnarray}  \label{3.2}
&&i\hbar\frac{\partial\Phi}{\partial t}=H_{({\rm 1st})}\Phi\,,\nonumber\\
H_{({\rm 1st})}&=&\sum_{(p,k)}\left(-\frac{\hbar^2}{2m_p}\right)
 \left[\nabla^{(p,k)}-\frac{i}{\hbar c}
 {\cal A}_{(p,k)}({\bf r}_1^{(1)},\ldots,{\bf r}_{n_M}^{(M)})-\frac{i}{\hbar c}
 e_p{\bf A}^{\rm ex}({\bf r}_k^{(p)},t)\right]^2\nonumber\\
& &+\sum_{(p,k)}\left(-\frac{\hbar}{2m_pc}\sigma_p\right)
   \epsilon^{ij}\nabla^{(p,k)}_i
   \left\{e_pA_j^{\rm ex}({\bf r}_k^{(p)})
   +{\cal A}_{(p,k)j}({\bf r}_1^{(1)},\ldots,{\bf r}_{n_M}^{(M)})\right\}\,,
\end{eqnarray}
since $\beta_{pp'}=\beta_{p'p}$ and we have, thanks to Eqs. (\ref{2.36}) and
(\ref{2.6}),
\begin{equation}
\epsilon^{ij}\nabla^{(p,k)}_i{\cal A}_{(p,k)j}=-\sum_{(p',k')\neq(p,k)}
  \beta_{pp'}\delta^2({\bf r}_k^{(p)}-{\bf r}_{k'}^{(p')})\,.
\end{equation}

Notice that $H_{({\rm 1st})}$ in Eq. (\ref{3.2}) has the form of the
non-relativistic (many-body) Pauli Hamiltonian on the plane, with spins
$\sigma_p=+1$ or $-1$. There is a hidden supersymmetry in the system \cite{8,9}
which can be exploited to find the exact many-body ground state. But, in this
paper, we shall direct our
interest to the static solutions of the corresponding classical field theory.
That is, we consider $\Psi_p({\bf r},t)$ to be classical $c$-number
fields and $\Psi_p^\dagger({\bf r},t)$
the corresponding complex conjugates $\Psi_p^*({\bf r},t)$.
The Hamiltonian is as in
Eq. (\ref{2.7}) (with the potential $U$ given by Eq. (\ref{3.1})), but the
classical equations of motion do not include the operator ordering term in
Eq. (\ref{2.9}), i.e.,
\begin{eqnarray}  \label{3.4}
i\hbar\frac{\partial\Psi_p({\bf r},t)}{\partial t}
&=&\sum_Iq_p^Ia^I_0({\bf r},t)\Psi_p({\bf r},t)
 -\frac{\hbar^2}{2m_p}{\bf D}^2\Psi_p({\bf r},t)
 -\frac{\hbar}{2m_pc}\sigma_pe_pB^{\rm ex}({\bf r})\Psi_p({\bf r},t)\nonumber\\
& &\hspace{-3mm}+\sum_{p'}\left(\frac{\hbar}{2m_pc}\sigma_p
     +\frac{\hbar}{2m_p'c}\sigma_{p'}\right)\beta_{pp'}
     \Psi_{p'}^*({\bf r},t)\Psi_{p'}({\bf r},t)\Psi_p({\bf r},t)\,,
\end{eqnarray}
where the C-S gauge fields $a^I_\mu({\bf r},t)$ are of course supposed to
satisfy Eqs. (\ref{2.5}) and (\ref{2.12}). As we will show, this classical
field system under suitable restrictions admits a class of interesting,
vortex-type, static solutions, carrying non-trivial characteristics endowed
upon them by the C-S interactions. Our work generalizes Refs.
\cite{11} and \cite{15}, where the case of a single matter field with
$B^{\rm ex}=0$ (Ref. \cite{11}) or $B^{\rm ex}\neq0$ (Ref.
\cite{15}) was analyzed. It is conceivable that the solitonlike solutions
discussed here may have significant physical roles as regards the nature of
various topological fluids within the effective field theory approach
\cite{3,4}.

To proceed, note that the choice of the potential as in Eq. (\ref{3.1}) allows
us to write the static energy functional in the form
\begin{eqnarray}
&&E=\int\! d^2\!{\bf r}\,
\sum_p\frac{\hbar^2}{2m_p}\left|D_1\Psi_p+i\sigma_pD_2\Psi_p\right|^2\,,\\
&&(D_i\Psi_p\equiv\left(\nabla_i-\frac{i}{\hbar c}\sum_Iq_p^Ia_i^I
   -\frac{i}{\hbar c}e_pA^{\rm ex}_i\right)\Psi_p, \hspace{8mm} i=1,2)\nonumber
\end{eqnarray}
dropping unimportant surface terms. This is an immediate consequence of the
identity
\begin{eqnarray}   \label{3.6}
|{\bf D}\Psi_p|^2&=&|(D_1+i\sigma_pD_2)\Psi_p|^2
     +\frac{1}{\hbar c}\sigma_p(e_pB^{\rm ex}
     +\sum_Iq_p^Ib^I)|\Psi_p|^2\nonumber\\
      && -i\sigma_p\epsilon^{ij}\nabla_i(\Psi_p^*D_j\Psi_p)
\end{eqnarray}
and the relation (\ref{2.4}). Hence any configuration satisfying the {\it
self-duality equations\/}
\begin{mathletters}
\begin{eqnarray}
&&D_1\Psi_p=-i\sigma_p D_2\Psi_p\,,\hspace{4mm}
      (\mbox{$\sigma_p=+1$ or $-1$})                  \label{3.7a}\\
&&\kappa_I\epsilon^{ij}\nabla_ia^I_j=-\sum_pq_p^I\Psi_p^*\Psi_p  \label{3.7b}
\end{eqnarray}
\end{mathletters}
will have the lowest possible energy, i.e., $E=0$. A solution of these
equations should solve the classical field equations (\ref{3.4})
automatically, but there is of course no guarantee that one can always find a
non-trivial solution to Eqs. (\ref{3.7a}) and (\ref{3.7b}). Suppose that a
non-trivial solution exists. Then we may conveniently write (i.e., work in the
Coulomb gauge)
\begin{equation}   \label{3.8}
a^I_i({\bf r})=-\epsilon^{ij}\nabla_jU^I({\bf r})\,,\hspace{6mm}
A_i^{\rm ex}({\bf r})=-\epsilon^{ij}\nabla_jV^{\rm ex}({\bf r})\,,\hspace{6mm}
\end{equation}
and introduce the functions $f_p({\bf r})$ ($p=1,\ldots,M$) by
\begin{equation}  \label{3.9}
\Psi_p({\bf r})=e^{-\frac{1}{\hbar c}\sigma_p\{\sum_Iq_p^IU^I({\bf r})
                         +e_pV^{\rm ex}({\bf r})\}}f_p({\bf r})\,.
\end{equation}
For the functions $f_p({\bf r})$, Eq. (\ref{3.7a}) now implies that
\begin{equation}
(\nabla_1+i\sigma_p\nabla_2)f_p({\bf r})=0\,,
\end{equation}
and therefore $f_p({\bf r})$ should be restricted to an entire function of
$z_{(\sigma_p)}\equiv x+i\sigma_py$,
viz., $f_p=f_p(z_{(\sigma_p)})$. Clearly, the function $\Psi_p({\bf r})$
may vanish only at the zeros of $f_p$ and let these zeros be at
$({\bf R}_1^{(p)},\ldots,{\bf R}_{n_p}^{(p)})$. Also, from Eq. (\ref{3.9}),
we have
\begin{equation}   \label{3.11}
\sum_Iq_p^IU^I({\bf r})+e_pV^{\rm ex}({\bf r})=-\sigma_p\frac{\hbar c}{2}
  \ln\left[|\Psi_p({\bf r})|^2/|f_p(z_{(\sigma_p)})|^2\right]\,,
\end{equation}
and combining these with Eq. (\ref{3.7b}) then yields the equations
\begin{equation}   \label{3.12}
\nabla^2\ln|\Psi_p|^2
=\sigma_p\frac{2}{\hbar c}\left\{\sum_{p'}\beta_{pp'}|\Psi_{p'}|^2
-e_pB^{\rm ex}\right\}+4\pi\sum_{r=1}^{n_p}\delta^2({\bf r}-{\bf R}_r^{(p)})\,,
       \hspace{3mm} (p=1,\ldots,M)
\end{equation}
where we have used the definition (\ref{2.15}) and the relation
$\nabla^2\ln|f_p|^2=4\pi\sum_{r=1}^{n_p}\delta^2({\bf r}-{\bf R}_r^{(p)})$.

Our problem is now reduced to the study of the coupled nonlinear equations
(\ref{3.12}), satisfied by the matter densities $|\Psi_p({\bf r})|^2$. Note
that, in
connection with the C-S interactions, only the parameters $\beta_{pp'}$
enter Eq. (\ref{3.12}); this is natural in view of our general observation
made in Sec. II. While the equations of this form appear very frequently,
say, in the study of various integrable models, there is no systematic
mathematical method for constructing the solutions for generic
$(\beta_{pp'})$, $M\ge2$. Nonetheless, the solution space is expected to have a
rich structure. With that in mind, we shall below study in some detail the
nature of solutions allowed when there are just two independent matter fields,
i.e. $M=2$, and the external magnetic field $B^{\rm ex}$ is zero or at most
uniform. Assuming this, we now write $(\Psi_1,\Psi_2)\equiv(\phi,\chi)$
and present Eq. (\ref{3.12}) in the form
\begin{eqnarray}  \label{3.13}
\nabla^2\ln|\phi|^2&=&\sigma_1(\bar\beta_{11}|\phi|^2
      +\bar\beta_{12}|\chi|^2-\bar{e}_1B^{\rm ex})
      +4\pi\sum_{r=1}^{n_1}\delta^2({\bf r}-{\bf R}_r)\,,\nonumber\\
\nabla^2\ln|\chi|^2&=&\sigma_2(\bar\beta_{12}|\phi|^2
      +\bar\beta_{22}|\chi|^2-\bar{e}_2B^{\rm ex})
      +4\pi\sum_{r=1}^{n_2}\delta^2({\bf r}-{\bf R}'_r)\,,
\end{eqnarray}
where we have denoted $\frac{2}{\hbar c}\beta_{pp'}=\bar\beta_{pp'}$ and
$\frac{2}{\hbar c}e_p=\bar{e}_p$.
Here it is not
difficult to guess that the system under non-zero $B^{\rm ex}$ behaves
differently from that for $B^{\rm ex}=0$. Also, a rather different behavior is
expected when the condition $\bar\beta_{11}\bar\beta_{22}=\bar\beta_{12}^2$
is satisfied. [See Eq.
(\ref{2.41}).] Hence, for the respective cases decided by these factors, we
will separately look for the solutions to Eq. (\ref{3.13}) below.

\subsection{The case with $B^{\rm ex}=0$ and
             $\Delta\equiv\bar\beta_{11}\bar\beta_{22}-\bar\beta_{12}^2\neq0$}
In this case, Eq. (\ref{3.13}) may be written in the form\footnote{Here and
henceforth, we will often omit the $\delta$-function terms in the equation,
which are significant only at the zeros of $\phi$ or $\chi$.}
\begin{equation}  \label{3.14}
\nabla^2\left(\begin{array}{l}\ln|\phi|^2\\ \ln|\chi|^2\end{array}\right)
 = -K\left(\begin{array}{l}|\phi|^2\\ |\chi|^2\end{array}\right),
\end{equation}
where
\begin{equation} \label{3.15}
K=\left(\begin{array}{ll}-\sigma_1\bar\beta_{11} &\ \ -\sigma_1\bar\beta_{12}\\
 -\sigma_2\bar\beta_{12} &\ \ -\sigma_2\bar\beta_{22}\end{array}\right)
 \equiv\left(\begin{array}{ll}K_{11} &\ \ K_{12}\\
                         K_{21} &\ \ K_{22}\end{array}\right).
\end{equation}
This is a Toda-type equation \cite{16} and we are here interested in the
regular solutions with $Q_\phi\equiv\int\! d^2\!{\bf r}\, |\phi|^2<\infty$ and
$Q_\chi\equiv\int\! d^2\!{\bf r}\, |\chi|^2<\infty$.
As we shall see below, the characters of
this equation depend very much on the properties of the matrix $K$.

The above system is integrable if $K$ has certain specific forms. To discuss
this case, we rescale the matter fields as $|\phi|^2=|\widetilde\phi|^2/a$ and
$=|\widetilde\chi|^2/b$ ($a$, $b$: positive constants) so that Eq. (\ref{3.14})
may assume the form
\begin{equation}   \label{3.16}
\nabla^2\left(\begin{array}{l}\ln|\widetilde\phi|^2\\
              \ln|\widetilde\chi|^2\end{array}\right)
 = -\widetilde K\left(\begin{array}{l}|\widetilde\phi|^2\\
                     |\widetilde\chi|^2\end{array}\right),
\hspace{5mm}
\widetilde K=\left(
  \begin{array}{ll}-\sigma_1\bar\beta_{11}/a &\ \ -\sigma_1\bar\beta_{12}/b\\
  -\sigma_2\bar\beta_{12}/a &\ \ -\sigma_2\bar\beta_{22}/b\end{array}\right).
\end{equation}
This equation will be integrable \cite{17} if the matrix
$\widetilde K$ is identified
with one of the Cartan matrices of the classical Lie algebra. We here recall
following rank-two Cartan matrices
\begin{equation}
A_2:\ \left(\begin{array}{rr}2 &\ \ -1\\ -1 &\ \ 2\end{array}\right),\qquad
B_2:\ \left(\begin{array}{rr}2 &\ \ -2\\ -1 &\ \ 2\end{array}\right),\qquad
G_2:\ \left(\begin{array}{rr}2 &\ \ -1\\ -3 &\ \ 2\end{array}\right).
\end{equation}
Comparing $\widetilde K$ with these expressions, we conclude that our system is
integrable if $\sigma_1=\sigma_2=\sigma(=\pm 1)$ and $K$ belongs to one
of the following one-parameter families:
\begin{equation}  \label{3.18}
K= a\left(\begin{array}{rr}-2 &\ \ 1\\ 1 &\ \ -2\end{array}\right),\ \
 a\left(\begin{array}{rr}-2 &\ \ 1\\ 1 &\ \ -1\end{array}\right),\ \
\mbox{or\ \ }a\left(\begin{array}{rr}-2 &\ \ 3\\ 3 &\ \ -6\end{array}\right).
\end{equation}
[Note that, in Eq. (\ref{3.18}), the constant $b$ appeared above has been set
to $a$, $\frac{1}{2}a$ or $\frac{1}{3}a$.] These special cases are known to be
also relevant for non-abelian self-dual C-S systems first discussed
in Ref. \cite{12}. We here have
$\Delta\equiv\bar\beta_{11}\bar\beta_{22}-\bar\beta_{12}^2>0$ and
hence, according to Eq. (\ref{2.39}), they all belong to the case with
$\kappa_1\kappa_2>0$. A constructive method for the corresponding exact
solutions is
described in Ref. \cite{13}. An interesting property of the solutions when
$\widetilde K$ is equal to the Cartan matrix of $A_2$ (and probably $B_2$
and $G_2$ as
well) is that the charges $Q_\phi$, $Q_\chi$ are quantized in such a way that
the fluxes (see Eq. (\ref{2.4}))
\begin{eqnarray} \label{3.19}
\Phi_\phi&\equiv &\int\! d^2\!{\bf r}\,
    \epsilon^{ij}\nabla_i(q_1^1a_j^1+q_1^2a_j^2)
=-(\beta_{11}Q_\phi+\beta_{12}Q_\chi)\,,\nonumber\\
    \Phi_\chi&\equiv &\int\! d^2\!{\bf r}\,
    \epsilon^{ij}\nabla_i(q_2^1a_j^1+q_2^2a_j^2)
=-(\beta_{21}Q_\phi+\beta_{22}Q_\chi)
\end{eqnarray}
may become integer multiples of $2\pi\hbar c$.

For a generic matrix $K$, Eq. (\ref{3.14}) has been studied little so far. For
example, there is no known criterion for the existence of a regular solution
to the equation. Therefore, we will first try to narrow down the range of the
parameters $\bar\beta_{pp'}$ (for given $\sigma_1$, $\sigma_2$)
in which a regular solution
with finite charges might be available, and then give some explicit solutions
for certain specific cases. First of all, we recall that a solution to Eqs.
(\ref{3.7a}) and (\ref{3.7b}), if exists, should have zero energy. Hence, in
view of Eq. (\ref{2.7}), there will be no regular self-dual solution if the
potential $U$ is positive definite. The potential is, from Eq. (\ref{3.1}),
given by (here, it suffices to set $m_1=m_2=m$ since the self-duality
equations are independent of the masses)
\begin{equation}  \label{3.20}
\frac{4m}{\hbar^2}U=\sigma_1\bar\beta_{11}|\phi|^4
      +(\sigma_1+\sigma_2)\bar\beta_{12}|\phi|^2|\chi|^2
                    +\sigma_2\bar\beta_{22}|\chi|^4\,.
\end{equation}
When both $\kappa_1$ and $\kappa_2$ are positive,
we have $\bar\beta_{11}>0$, $\bar\beta_{22}>0$ and
$\bar\beta_{11}\bar\beta_{22}-\bar\beta_{12}^2>0$ and so conclude that $U$ is
positive-definite if $\sigma_1=\sigma_2=1$. With $\kappa_1<0$ and
$\kappa_2<0$, $U$ will be positive-definite if $\sigma_1=\sigma_2=-1$.
On the other hand, with $\kappa_1>0$, and $\kappa_2<0$ (and
therefore $\bar\beta_{11}\bar\beta_{22}-\bar\beta_{12}^2<0$), we will have
positive-definite potential if all the coefficients appearing in the right
hand side of Eq. (\ref{3.20}) are positive, viz., no self-dual solution if
$K_{11}<0$, $K_{22}<0$ and (for the case of $\sigma_1\sigma_2=1$) $K_{12}<0$.
The parameter range is also restricted by the fact that if the Laplacian of a
function $f$ is positive for large $|{\bf r}|$, $f$ is asymptotically
increasing. We may apply this with $f=-\ln|\phi|^2$ or $-\ln|\chi|^2$,
noting that $|\phi|^2$ and
$|\chi|^2$ should approach zero as $|{\bf r}|\rightarrow\infty$
(and hence the functions $-\ln|\phi|^2$ and $-\ln|\chi|^2$ increase
indefinitely) to obtain a configuration with finite charges.
Then, in view of Eq. (\ref{3.14}), this
cannot be realized if the matrix $K$ is strictly negative; hence, no solution
if $K_{pp'}(=-\sigma_p\bar\beta_{pp'})<0$ for every $p$, $p'$.

For further restrictions on the parameters, it is useful to note that Eq.
(\ref{3.14}), now including the $\delta$-function terms, can be cast as
\begin{eqnarray} \label{3.21}
\nabla^2\ln\left(\frac{\left\{|\chi|^2/\prod_{r=1}^{n_2}
         |z-Z'_r|\right\}^{K_{12}}}{\left\{|\phi|^2/
 \prod_{r=1}^{n_1}|z-Z_r|\right\}^{K_{22}}}\right)
&=&(\det K)|\phi|^2\,,\nonumber\\
\nabla^2\ln\left(\frac{\left\{|\phi|^2/\prod_{r=1}^{n_1}
         |z-Z_r|\right\}^{K_{21}}}{\left\{|\chi|^2/
 \prod_{r=1}^{n_2}|z-Z'_r|\right\}^{K_{11}}}\right)&=&(\det K)|\chi|^2\,.
\end{eqnarray}
If
$\det K=\sigma_1\sigma_2\{\bar\beta_{11}\bar\beta_{22}-(\bar\beta_{12})^2\}$
is positive, these relations show that both
$\displaystyle
 \frac{\left\{|\chi|^2/\prod_{r=1}^{n_2}
 |z-Z'_r|\right\}^{K_{12}}}{\left\{|\phi|^2/
 \prod_{r=1}^{n_1}|z-Z_r|\right\}^{K_{22}}}$ and $\displaystyle
 \frac{\left\{|\phi|^2/\prod_{r=1}^{n_1}|z-Z_r|
 \right\}^{K_{21}}}{\left\{|\chi|^2/
 \prod_{r=1}^{n_2}|z-Z'_r|\right\}^{K_{11}}}$ should be asymptotically
increasing; but, this cannot be the case if $K_{12}>0$ and $K_{22}<0$ or if
$K_{21}>0$ and $K_{11}<0$. With $\det K <0$, we encounter a similar
inconsistency if $K_{12}<0$ and $K_{22}>0$ or if $K_{21}<0$ and $K_{11}>0$.
Based on this discussion, we conclude that {\it no\/} solution can be found
in the following parameter ranges:
\begin{enumerate}
\item[(i)] ($\sigma_1=\sigma_2=+1$,
         $\bar\beta_{12}<0$) or ($\sigma_1=-\sigma_2,$ $\bar\beta_{12}>0$),
         with $\kappa_1>0$ and $\kappa_2>0$,
\item[(ii)] ($\sigma_1=\sigma_2=-1$,
         $\bar\beta_{12}>0$) or ($\sigma_1=-\sigma_2,$ $\bar\beta_{12}<0$),
         with $\kappa_1<0$ and $\kappa_2<0$,
\item[(iii)]  $\sigma_1=\sigma_2=\sigma(=\pm 1)$,
         $K_{12}=-\sigma\bar\beta_{12}<0$, and at least one between
         $K_{11}(=-\sigma\bar\beta_{11}$) and $K_{22}(=-\sigma\bar\beta_{22}$)
         is positive, with $\kappa_1\kappa_2<0$,
\item[(iv)] $\sigma_1\sigma_2=-1$,
         $\bar\beta_{12}\bar\beta_{22}>0$ or $\bar\beta_{12}\bar\beta_{11}>0$,
         with $\kappa_1\kappa_2<0$.
\end{enumerate}

We have summarized our findings in Fig. 2. The shaded regions are those
excluded on the basis of the above arguments, viz., there does not exist a
non-trivial solution satisfying the self-duality equations if the parameters
lie in these regions. Note that exactly a half of the parameter space is
excluded.

Let us now discuss some explicit solutions to the self-duality equations in
question. A glance at Eq. (\ref{3.14}) shows that, with the ansatz
$|\chi({\bf r})|^2=\gamma|\phi({\bf r})|^2$ ($\gamma$: a positive constant), it
can be reduced to a single equation for $|\phi|^2$ as long as $\gamma$ is
chosen suitably. Indeed, as we make the choice
\begin{equation}
\gamma=\frac{K_{11}-K_{21}}{K_{22}-K_{12}}\,,
\end{equation}
what follows from Eq. (\ref{3.14}) is
just the Liouville equation for $|\phi|^2$:
\begin{equation}   \label{3.23}
\nabla^2\ln|\phi|^2=\frac{(\det K)}{K_{12}-K_{22}}|\phi|^2\,.
\end{equation}
To have non-trivial solutions, the coefficient of $|\phi|^2$ on the right hand
side of Eq. (\ref{3.23}) should be negative as well as $\gamma>0$. For these
Liouville-type solutions, the fluxes are quantized, i.e.,
$\Phi_\phi=\Phi_\chi=(\mbox{integer multiples of $2\pi\hbar c$})$. We here note
that, due to the restrictions mentioned above, the parameter range allowing
these Liouville-type solutions does not cover the entire unshaded region in
Fig. 2. One may then ask
\begin{enumerate}
\item[(i)]When the Liouville-type solutions exist, are there additional
solutions distinct from the Liouville-type? If so, should they have quantized
flux values always?
\item[(ii)]When the parameters are such that no Liouville-type solution
exists, will there be some solutions to Eq. (\ref{3.14}) after all?
\end{enumerate}
These issues are discussed below.

On the question (i), the existence of more general type solutions is confirmed
by the index theorem, and our numerical study further shows that there are also
solutions with nonquantized flux values. The result of the index theorem
analysis (see Appendix B for details) is as follows: the number of free
parameters in the general solution, with the field $\phi$ ($\chi$) having
vorticity
$n_1$ ($n_2$) and the asymptotic behavior
$\sim\frac{1}{r^{\alpha_1}}$ ($\sim\frac{1}{r^{\alpha_2}}$), is equal to
$2(n_1+\hat\alpha_1)+2(n_2+\hat\alpha_2)$,
where $\hat\alpha$ denotes the largest integer
less than $\alpha$. Here the `vorticity' is equal to the total number of zeros
in a specific matter field. By applying this to the case when the
Liouville-type solutions appropriate
to the values $n_1=n_2=n$ and $\alpha_1=\alpha_2=n+2$ are allowed, we
immediately see that the corresponding general solution should contain
$8(n+1)$ free parameters. But this is precisely twice the number of free
parameters entering the Liouville-type solutions \cite{18}. A plausible
conjecture is that, in general solution, we no longer have the restriction
(satisfied by all Liouville-type solutions) that the zeros of $\phi$ are also
the zeros of $\chi$. Furthermore, we found numerically
that there exist also solutions with the asymptotic behaviors not given by
integral power falloff, i.e., $\alpha_1$ and $\alpha_2$ above need not be
integers but
vary continuously. For an example, see the next-to-next paragraph. In view
of the relations $\Phi_p=\pm2\pi\hbar c(n_1+\alpha_1)$ and
$\Phi_\chi=\pm2\pi\hbar
c(n_2+\alpha_2)$, these solutions will then have {\it nonquantized\/}
fluxes\footnote{Recently some authors \cite{19} argued that the flux (or
charge) for the solutions of the Liouville equation is quantized because of
the inversion symmetry. This is misleading, however. Rather, we might as well
say that the inversion transformation generates {\it non-singular\/} solutions
because the charges happens to be quantized (and, correspondingly, the field
has integral power falloff asymptotically). No useful information is gained by
considering the inversion symmetry in our case.}. [In this regard, see also
the comment immediately after Eq. (\ref{3.36})].

In the parameter range where no Liouville-type solution exists, analytical
means are not available at present and we resorted to numerical analysis
(assuming the rotationally symmetric form). For some values of parameters we
succeeded in finding solutions while, for other values, no acceptable
solution could be found. It seems that solutions exist in a large portion of
this parameter range also, but the precise criterion on the parameters to have
solutions is not clear yet. We also note that if a particular
non-Liouville-type solution, say
$\pmatrix{|\widetilde\phi(z,z^*)|\cr |\widetilde\chi(z,z^*)|}$, is known,
a family of new
solutions may be obtained by considering its conformal transformation
\cite{12}, i.e.,
$\pmatrix{|\varphi'(z)||\widetilde\phi(\varphi(z),\varphi^*(z^*))|\cr
          |\varphi'(z)||\widetilde\chi(\varphi(z),\varphi^*(z^*))|}$
for a polynomial function $\varphi(z)$.

Now, as a specific example, we will discuss solutions in the self-dual system
with $\beta_{11}=\beta_{22}=0$ but nonzero $\beta_{12}$ (i.e. keep only mutual
statistical interaction). This can be realized by the choice
\begin{equation}  \label{3.24}
\kappa_1=-\kappa_2\equiv\kappa\,,\qquad q_1^1
    =q_1^2\equiv\frac{1}{\sqrt{2}}q\,,\qquad q_2^1=-q_2^2\equiv q'\,,
\end{equation}
and then $\beta_{12}=\frac{qq'}{\kappa}$. The self-duality equations now read
\begin{eqnarray}  \label{3.25}
\nabla^2\ln|\phi|^2&=&-\frac{2}{\hbar
c}\frac{qq'}{\kappa}|\chi|^2\,,\nonumber\\
\nabla^2\ln|\chi|^2&=&-\frac{2}{\hbar c}\frac{qq'}{\kappa}|\phi|^2\,,
\end{eqnarray}
choosing $\sigma_1=\sigma_2=-1$ so that solutions may exist.
Note that if the theory
is rewritten in terms of the fields $v_\mu^{(1)}$ and $v_\mu^{(2)}$ defined by
Eq. (\ref{2.42}), this system is described by the Lagrangian density
\begin{eqnarray}  \label{3.26}
{\cal L}&=&\kappa\epsilon^{\mu\nu\lambda}v_\mu^{(1)}
 \partial_\nu v_\lambda^{(2)}+i\hbar\phi^*
 \left(\frac{\partial}{\partial t}-\frac{i}{\hbar}qv_0^{(1)}\right)\phi
 +i\hbar\chi^*\left(\frac{\partial}{\partial t}
 -\frac{i}{\hbar}q'v_0^{(2)}\right)\chi\nonumber\\
 & &-\frac{\hbar^2}{2m_1}
        \left|\left(\nabla-\frac{i}{\hbar c}q{\bf v}^{(1)}\right)\phi\right|^2
        -\frac{\hbar^2}{2m_2}\left|\left(\nabla-\frac{i}{\hbar c}
        q'{\bf v}^{(2)}\right)\chi\right|^2\nonumber\\
&&+\left(\frac{\hbar}{2m_1c}
        +\frac{\hbar}{2m_2c}\right)\frac{qq'}{\kappa}|\phi|^2|\chi|^2\,.
\end{eqnarray}
The equations (\ref{3.25}) clearly admit the Liouville-type solutions, which
are based on the ansatz $|\phi({\bf r})|=|\chi({\bf r})|$.
In addition, we have found
some non-Liouville-type solutions numerically, assuming that the fields
$\phi({\bf r})$ and $\chi({\bf r})$ are functions of $r$ only. These are
shown in Fig.
3. In Fig. 3(a), a plot is given for a solution corresponding to $n_1=n_2=0$
(i.e. zero vorticity for both $\phi$ and $\chi$), the asymptotic behavior of
which is determined as $|\phi|\sim r^{-1.34}$ and $|\chi|\sim r^{-3.95}$.
Evidently, the fluxes are not quantized for this solution. Another plot now
for a solution corresponding to $n_1=0$ and $n_2=1$ is given in Fig. 3(b).

Finally note that the charges $(Q_\phi,Q_\chi)$ of a vortex soliton with given
fluxes $(\Phi_\phi,\Phi_\chi)$ are determined by Eq. (\ref{3.19}), and they
will be
in general non-zero. This prompts one to make a conjecture that the vortices
described above experience fractional and also mutual statistical interactions
analogous to those experienced by the elementary quanta in the theory. The
most direct approach to settle this question is to derive the effective
Lagrangian which is relevant to the slow dynamics of these vortices \cite{20}.
But, in our case, this program is nontrivial especially because a multi-vortex
type solution cannot be interpreted in an unambiguous way as representing a
collection of some elementary vortices here. So it remains as an open
problem.

\subsection{The case with $B^{\rm ex}=0$ and
                    $\Delta=\bar\beta_{11}\bar\beta_{22}-\bar\beta_{12}^2=0$}
If $\bar\beta_{11}\bar\beta_{22}$ is equal to $\bar\beta_{12}^2$, two C-S
gauge fields are no
longer independent and hence we are led to a self-dual system with two matter
fields coupled to a single C-S gauge field. In fact, we may now write
$D_i\phi=(\nabla_i-\frac{i}{\hbar c}q'_1a_i)\phi$,
$D_i\chi=(\nabla_i-\frac{i}{\hbar c}q'_2a_i)\chi$, and
$\bar\beta_{ij}=\frac{2}{\hbar c}\frac{q'_iq'_j}{\kappa}$ ($i,j=1,2$).
Here we find it convenient to write Eq. (\ref{3.13}) using the couplings
$q_i\equiv\sqrt{\frac{2}{\hbar c |\kappa|}}q'_i$ rather than
$\bar\beta_{ij}$, viz.,
\begin{eqnarray}  \label{3.41}
\nabla^2\ln|\phi|^2&=&-\sigma_1q_1(q_1|\phi|^2+q_2|\chi|^2)\,,\nonumber\\
\nabla^2\ln|\chi|^2&=&-\sigma_2q_2(q_1|\phi|^2+q_2|\chi|^2)\,,
\end{eqnarray}
where we have chosen $\kappa<0$ (with no loss of generality). Evidently,
if $q_1$
and $q_2$ are of the same sign (i.e. $q_1q_2>0$), these equations will admit a
bounded solution only for $\sigma_1=\sigma_2=+1$. For $q_1q_2<0$, on
the other hand, one can have a solution only with $\sigma_1\sigma_2=-1$.
To show this, suppose that
$\sigma_1=\sigma_2=1$. Then, from Eq. (\ref{3.41}) (with the suppressed
$\delta$-function terms put in), we have
\begin{equation}
\nabla^2\ln|\phi|^2-\frac{q_1}{q_2}\nabla^2\ln|\chi|^2
 =4\pi\sum_{r=1}^{n_1}\delta^2({\bf r}-{\bf R}_r)-4\pi\left(
  \frac{q_1}{q_2}\right)\sum_{r=1}^{n_2}\delta^2({\bf r}-{\bf R}'_r)\,,
\end{equation}
and hence
\begin{equation} \label{3.43}
\frac{\phi}{\chi^{q_1/q_2}}=
 C\frac{\prod_{r=1}^{n_1}(z-Z_r)}{\prod_{r=1}^{n_2}(z-Z'_r)^{q_1/q_2}}\,,
  \hspace{4mm}\mbox{($C$: a complex number)}.
\end{equation}
But, with $q_1q_2<0$, this is impossible: the left hand side of Eq.
(\ref{3.43}) vanishes asymptotically, while its left hand side clearly does
not. The situation is analogous for $\sigma_1=\sigma_2=-1$. Hence, the choice
$\sigma_1\sigma_2=-1$ is appropriate in the case of $q_1q_2<0$.

Incidentally, the above consideration is sufficient to show that there is no
solution with finite charge if the matrix $\widetilde{K}$, defined in Eq.
(\ref{3.16}), is equal to the $2\times2$ affine Cartan matrix
$\left(\begin{array}{rr}2&-2\\-2&2\end{array}\right)$ or
$\left(\begin{array}{rr}2&-4\\-1&2\end{array}\right)$. [This corresponds to the
case for which Eq. (\ref{3.16}) can be reduced to the sinh-Gordon or
Bullough-Dodd equation, as noted in Ref. \cite{12}.] Indeed, for these
particular forms for $\widetilde K$, we have $\sigma_1\sigma_2>0$ and then,
by the above consideration, a bounded solution may be possible only
with $\sigma_1=\sigma_2=+1$. So choose $\sigma_1=\sigma_2=+1$
and then we find the matrix $(\bar\beta_{pp'})$ described by the
one-parameter family of singular matrices,
$a\left(\begin{array}{rr}-2&2\\2&-2\end{array}\right)$ or
$a\left(\begin{array}{rr}-2&1\\1&-\frac{1}{2}\end{array}\right)$. When
translated into the above notation, this matrix form leads to $q_1q_2<0$ (i.e.
the wrong sign); hence, there should not be any bounded solution.

Let us now study soliton solutions for some representative cases. First we
consider the system with $q_1=q_2\equiv q$ (and $\sigma_1=\sigma_2=+1$, of
course), so that Eq. (\ref{3.41}) may assume the form
\begin{eqnarray}   \label{3.44}
\nabla^2\ln|\phi|^2&=&-q^2(|\phi|^2+|\chi|^2)\,,\nonumber\\
\nabla^2\ln|\chi|^2&=&-q^2(|\phi|^2+|\chi|^2)\,.
\end{eqnarray}
The corresponding system has additional global $SU(2)$ symmetry and may be
viewed as a nonrelativistic version of a special self-dual model considered by
Kim \cite{21}. By exploiting this global $SU(2)$ symmetry, a series of
exact solutions, which are different from the Liouville-type solutions
(obtained under the ansatz $|\chi|^2=\gamma|\phi|^2$), can be obtained.
Specifically, we found that the coupled equations in Eq. (\ref{3.44})
have also the solutions of the following type
\begin{equation}  \label{3.45}
\left(\begin{array}{c}\phi\\ \chi\end{array}\right)
=\frac{\sqrt{12}(P(z)Q'(z)-Q(z)P'(z))}{|q|(|P(z)|^2+|Q(z)|^2)^{3/2}}
  \left(\begin{array}{c}P(z)\\ Q(z)\end{array}\right)\,,
\end{equation}
where $P(z)$ and $Q(z)$ are arbitrary polynomials of $z$ under the restriction
that these two functions share no common zero. Note that this solution in Eq.
(\ref{3.45}) do not satisfy the Liouville equation and include no rotationally
symmetric configuration since $P(z)$ and $Q(z)$ have no common zero. The
fluxes $\Phi_\phi$ and $\Phi_\chi$ for these solutions are quantized as in the
Liouville-type solutions; in detail, for the solution (\ref{3.45}), we have
$\Phi_\phi=\Phi_\chi=2\pi\hbar c(3n_P+3n_Q)$ if $P(z)$ ($Q(z)$) is an
$n_P$ ($n_Q$)-th order polynomial. [Since the vorticities of $\phi$
and $\chi$ are equal to $n_1=2n_P+n_Q-1$ and $n_2=n_P+2n_Q-1$, this
may also be written as $\Phi_\phi=\Phi_\chi=2\pi\hbar c(n_1+n_2+2)$].
However, using the index theorem argument (see
Appendix B), we know that there must be solutions other than these two types.
We
are also not sure whether or not the fluxes are necessarily quantized for {\it
all\/} bounded solutions to Eq. (\ref{3.44}).

As another case, we choose $q_2/q_1=-2$ and $\sigma_1=-\sigma_2=+1$. Then,
according to the same procedure which led to Eq. (\ref{3.43}), we have
\begin{equation}
\frac{\chi^*}{\phi^2}=
  C^*\frac{\prod_{r=1}^{n_2}(z-Z'_r)}{\prod_{r=1}^{n_1}(z-Z_r)^2}\,,
  \hspace{4mm}\mbox{($C$: a complex number)}
\end{equation}
and inserting this into Eq. (\ref{3.41}) yields a single equation for $|\phi|$:
\begin{equation}  \label{3.47}
\nabla^2\ln|\phi|^2=-q_1^2\left\{|\phi|^2-2|C|^2
     \frac{\prod_{r=1}^{n_2}|z-Z'_r|^2}{\prod_{r=1}^{n_1}|z-Z_r|^4}|\phi|^4
     \right\}\,.
\end{equation}
If we here restrict ourselves to the special case
\begin{equation}
\frac{\chi^*}{\phi^2}=C^*\,,
\end{equation}
Eq. (\ref{3.47}) is simplified as
\begin{equation}
\nabla^2\ln|\phi|^2=-q_1^2|\phi|^2(1-2|C|^2|\phi|^2)\,.
\end{equation}
This is identical to the equation encountered in the relativistic self-dual
C-S Higgs system of Refs. \cite{22,23}. For the latter system, there are now
rigorous existence proofs \cite{24} for both topological and nontopological
soliton solutions. Only the nontopological ones are relevant in our case, for
the other class leads to infinite charge. In particular, a rotationally
symmetric nontopological soliton solution has the behaviors
\begin{equation}  \label{3.36}
|\phi({\bf r})|\sim\left\{\begin{array}{cl}
    r^n\ \ \ &\mbox{(for some nonnegative integer $n$), as $r\rightarrow0$}\\
    \frac{1}{r^\alpha}\ \ \
     &\mbox{(with $\alpha>n+2$), as $\rightarrow\infty$}\end{array}
  \right.
\end{equation}
and for this solutions we find the (nonquantized) fluxes
$\Phi_\phi=2\pi\hbar c(n+\alpha)$ and
$\Phi_\chi=-2\pi\hbar c(2n+2\alpha)$.
This is another evidence for our assertion that quantized flux
values are not to be expected generally.

\subsection{The case with $B^{{\rm ex}}\neq 0$}
As a uniform external magnetic field is turned on, a spontaneously broken
vacuum becomes possible and correspondingly we might then have nontrivial
solutions to Eq. (\ref{3.13}) in the form of {\it topological\/} solitons.
With a single matter field, an analogous phenomenon was noticed in Ref.
\cite{15}. Assuming $\Delta\equiv\bar\beta_{11}\bar\beta_{22}
-\bar\beta_{12}^2\neq0$ (the $\Delta=0$
case is considered later), the asymptotic values of $|\phi({\bf r})|$ and
$|\chi({\bf r})|$ for a topological soliton solution should be equal to
\begin{eqnarray}  \label{3.37}
v_\phi&=&\sqrt{B^{{\rm ex}}(\bar\beta_{22}\bar{e}_1
  -\bar\beta_{12}\bar{e}_2)/\Delta}\,,\nonumber\\
v_\chi&=&\sqrt{B^{{\rm ex}}(-\bar\beta_{12}\bar{e}_1
  +\bar\beta_{11}\bar{e}_2)/\Delta}\,.
\end{eqnarray}
Using these vacuum values, we may now rewrite the self-duality equations as
(cf. Eq. (\ref{3.14}))
\begin{equation}   \label{3.38}
\nabla^2\left(\begin{array}{c}\ln|\phi|^2\\ \ln|\chi|^2\end{array}\right)
 =-K\left(
 \begin{array}{c}|\phi|^2-v_\phi^2\\ |\chi|^2-v_\chi^2\end{array}\right)\,,
\end{equation}
with the same matrix $K$ as in Eq. (\ref{3.15}). In case a topological soliton
solution is allowed, it will be subject to the topological quantization
conditions of the form
\begin{eqnarray}
\int\! d^2\!{\bf r}\,\left[\epsilon^{ij}\nabla_i(q_1^1a_j^1+q_1^2a_j^2)
 +e_1B^{{\rm ex}}\right]
                             &=&2\pi\hbar cn_1\,,\nonumber\\
\int\! d^2\!{\bf r}\,\left[\epsilon^{ij}\nabla_i(q_2^1a_j^1+q_2^2a_j^2)
 +e_2B^{{\rm ex}}\right]
                             &=&2\pi\hbar cn_2\,,
\end{eqnarray}
where $n_1$ and $n_2$ are integers.

In view of Eq. (\ref{3.37}), one may hope to find a topological soliton
solution only when the parameters of the theory satisfy certain restrictions;
namely, for $\Delta>0$ (and hence $\kappa_1\kappa_2>0$), one must have
\begin{equation}  \label{3.40}
B^{{\rm ex}}(\bar\beta_{22}\bar{e}_1-\bar\beta_{12}\bar{e}_2)>0\,,\qquad
B^{{\rm ex}}(-\bar\beta_{12}\bar{e}_1+\bar\beta_11\bar{e}_2)>0\,,
\end{equation}
while, for $\Delta<0$ (and hence $\kappa_1\kappa_2<0$), the inequality
signs in Eq.
(\ref{3.40}) should be reversed. Aside from these, there must be some
conditions involving the elments of the matrix $K$ mainly. We can obtain such
conditions by studying Eq. (\ref{3.38}) in the asymptotic region. For this
asymptotic analysis, we set
\begin{equation}
f_\phi({\bf r})=\frac{1}{v_\phi^2}(v_\phi^2-|\phi({\bf r})|^2)\,,\hspace{4mm}
f_\chi({\bf r})=\frac{1}{v_\chi^2}(v_\chi^2-|\chi({\bf r})|^2)\,,
\end{equation}
and may instead study the linearized form of Eq. (\ref{3.38}), i.e.,
\begin{equation}   \label{3.42b}
\nabla^2\left(\begin{array}{c}f_\phi\\ f_\chi\end{array}\right)
 =-L\left(\begin{array}{c}f_\phi\\ f_\chi\end{array}\right)\,,\hspace{5mm}
L=\left(\begin{array}{cc}v_\phi^2K_{11}& v_\chi^2K_{12}\\
                         v_\phi^2K_{21}& v_\chi^2K_{22}\end{array}\right)\,.
\end{equation}
Now suppose that the $2\times2$ matrix $L$ can be diagonalized, i.e.,
$SLS^{-1}=\pmatrix{\lambda_1 & 0\cr 0& \lambda_2}$ for some non-singular
matrix $S$.  Then Eq. (\ref{3.42b}) can be cast as
\begin{equation}
\nabla^2\left(
 \begin{array}{c}\widetilde f_\phi\\ \widetilde f_\chi\end{array}\right)
 =-\left(\begin{array}{cc}\lambda_1&0\\ 0&\lambda_2\end{array}\right)
  \left(\begin{array}{c}\widetilde f_\phi\\
 \widetilde f_\chi\end{array}\right)\,,\hspace{5mm}
\left(\begin{array}{c}\widetilde f_\phi\\ \widetilde f_\chi\end{array}\right)
 \equiv S\left(\begin{array}{c}f_\phi\\ f_\chi\end{array}\right)\,.
\end{equation}
Here the eigenvalues $\lambda_1$ and $\lambda_2$, which have crucial
importance in
determining the asymptotic behaviors, are the roots of the secular equation
\begin{eqnarray}   \label{3.44b}
0&=&\det(L-\lambda I)\nonumber\\
&=&\lambda^2-(v_\phi^2K_{11}+v_\chi^2K_{22})\lambda+v_\phi^2v_\chi^2(\det K)\,.
\end{eqnarray}

Now take the case of $\kappa_1\kappa_2>0$ (and so $\Delta>0$). In this case,
it is easy
to show that Eq. (\ref{3.44b}) has necessarily two real roots. Then note that,
for an acceptable soliton solution, $f_\phi$ and $f_\chi$ above should approach
zero asymptotically in such a way that the resulting soliton may have finite
energy. This requires both roots to be negative (i.e. $\lambda_1<0$ and
$\lambda_2<0$), and, in view of Eq. (\ref{3.44b}), this translates into the
following conditions:
\begin{eqnarray}
&&v_\phi^2K_{11}+v_\chi^2K_{22}\equiv v_\phi^2\sigma_1\beta_{11}
  +v_\chi^2\sigma_2\beta_{22}<0\,,\nonumber\\
&&\det K=\sigma_1\sigma_2\Delta>0\,.
\end{eqnarray}
Based on these, we find that the appropriate choice for a nontrivial soliton
solution is
\begin{eqnarray}
\sigma_1=\sigma_2=-1
  &,\hspace{5mm}&\mbox{if $\kappa_1>0$ and $\kappa_2>0$}\nonumber\\
\sigma_1=\sigma_2=+1&,\hspace{5mm}&\mbox{if $\kappa_1<0$ and $\kappa_2<0$}\,.
\end{eqnarray}
A similar analysis may be repeated for the case of $\kappa_1\kappa_2<0$. For
the
latter case, however, the roots of Eq. (\ref{3.44b}) are not always real and
this introduces a certain uncertain feature in the analysis. Nevertheless, for
$\kappa_1\kappa_2<0$, we can make the following definite statement:{\it no\/}
solution exists with $\sigma_1\sigma_2=+1$.

In addition to the above, certain (plausible) conditions can also
be derived on the basis of the conjecture that a regular soliton
solution, if exists, is likely to fulfill the inequalities
\begin{equation}
|\phi({\bf r})|-v_\phi<0\,,\qquad |\chi({\bf r})|-v_\chi<0
\end{equation}
at least for sufficiently large $r$. Then, accepting this behavior, it is
possible to apply the same reasoning as in the case of $B^{{\rm ex}}=0$
(see the subsection A). For instance, Eq. (\ref{3.38}) will be inconsistent
with the assumed asymptotic behavior of $|\phi({\bf r})|$ and
$|\chi({\bf r})|$ if
the matrix $K$ is strictly positive; hence, no solution to Eq. (\ref{3.38}) if
$K_{pp'}=\sigma_p\bar\beta_{pp'}>0$ for every $p$, $p'$. Also,
by proceeding as in Eq. (\ref{3.21}), we expect that a
nontrivial solution may exist only under the conditions
\begin{eqnarray}
(\det K)\{n_1K_{22}-n_2K_{12}\}&<&0\,,\nonumber\\
(\det K)\{-n_1K_{21}+n_2K_{11}\}&<&0\,,
\end{eqnarray}
where $n_1$ ($n_2$) denotes the vorticity of the field $\phi$ ($\chi$).

We do not know of any analytic method developed to study the system in Eq.
(\ref{3.38}), even for some special $K$. We thus looked for numerical
solutions to Eq. (\ref{3.38}), while assuming the rotationally symmetric field
configurations. Here it should suffices to say that, at least for certain
choices of parameters (and vorticities) which satisfy the conditions given
above, we did confirm the existence of regular topological soliton
solutions. Note that, in view of the index theorem (see Appendix B), the
existence of a particular solution with vorticities $n_1$ and $n_2$ actually
imply the existence of a $2(n_1+n_2)$ parameter family of soliton solutions.

If $\Delta$ happens to vanish, we again have a self-dual system with two matter
fields coupled to a single C-S gauge field. Using the same notation as in the
subsection B, it will be possible to rewrite Eq. (\ref{3.38}) (with the choice
$\kappa<0$) as
\begin{eqnarray}
\nabla^2\ln|\phi|^2&=&-\sigma_1\left[q_1(q_1|\phi|^2+q_2|\chi|^2)
 -\bar e_1B^{{\rm ex}}\right]\,,\nonumber\\
\nabla^2\ln|\chi|^2&=&-\sigma_2\left[q_2(q_1|\phi|^2+q_2|\chi|^2)
 -\bar e_2B^{{\rm ex}}\right]\,.
\end{eqnarray}
A precondition to have a topological soliton solution is the existence of a
non-trivial (uniform) vacuum solution, and in the present case this will be
true only when the coupling parameters satisfy the relation
\begin{equation}  \label{3.44a}
\frac{e_1}{q_1}=\frac{e_2}{q_2}\,.
\end{equation}
Here a particularly simple case is obtained for $e_1=e_2=e$ and then, thanks
to Eq. (\ref{3.44a}), $q_1=q_2=q$. For this special case, the system has in
fact a global $SU(2)$ symmetry and this is also manifest in the self-duality
equations
\begin{eqnarray}  \label{3.45a}
\nabla^2\ln|\phi|^2&=&-\sigma_1\left[q^2(|\phi|^2+|\chi|^2)
 -\bar eB^{{\rm ex}}\right]\,,\nonumber\\
\nabla^2\ln|\chi|^2&=&-\sigma_2\left[q^2(|\phi|^2+|\chi|^2)
 -\bar eB^{{\rm ex}}\right]\,.
\end{eqnarray}
This system admits a topological soliton solution if we choose
$\sigma_1=\sigma_2=-1$ and $eB^{{\rm ex}}>0$.
Then, Eq. (\ref{3.45a}) becomes identical to the self-duality
equations found in the relativistic self-dual Ginzburg-Landau model with the
so-called semilocal symmetry \cite{25}. General solutions for this model are
discussed in those papers to which readers are referred.

\section{RELATIVISTIC GENERALIZATION}
Here we will introduce the relativistic self-dual $U(1)\times U(1)$
C-S system (with no external magnetic field, for simplicity) and then
study the static soliton solutions in the model. If one wishes, one may view
this investigation as a direct generalization of the model considered in
Ref. \cite{22}. Our model contains two complex scalar fields $\phi$ and $\chi$,
and is described by the Lagrangian density
\begin{equation}  \label{4.1}
{\cal L}=\frac{\kappa_1}{2}\epsilon^{\mu\nu\lambda}
     a_\mu^1\partial_\nu a_\lambda^1
     +\frac{\kappa_2}{2}\epsilon^{\mu\nu\lambda}a_\mu^2\partial_\nu a_\lambda^2
     -|D_\mu\phi|^2-|D_\mu\chi|^2-U(\phi,\chi)\,,
\end{equation}
where
\begin{eqnarray}
D_\mu\phi &\equiv&[\partial_\mu-i(q_1^1a_\mu^1+q_1^2a_\mu^2)]\phi\,,\nonumber\\
D_\mu\chi&\equiv &[\partial_\mu-i(q_2^1a_\mu^1+q_2^2a_\mu^2)]\chi\,,
\end{eqnarray}
and $U(\phi,\chi)$ is to be chosen shortly. [In this section, we set
$c=\hbar=1$.] The Gauss laws read
\begin{eqnarray}  \label{4.3}
b^1 &\equiv &\epsilon^{ij}\nabla_ia_j^1
 =-\frac{1}{\kappa_1}(q_1^1J_\phi^0+q_2^1J_\chi^0)\,,\nonumber\\
b^2 &\equiv &\epsilon^{ij}\nabla_ia_j^2
 =-\frac{1}{\kappa_2}(q_1^2J_\phi^0+q_2^2J_\chi^0)
\end{eqnarray}
with the currents $J_\phi^\mu\equiv -i[\phi^*D^\mu\phi-(D^\mu\phi)^*\phi]$ and
$J_\chi^\mu\equiv -i[\chi^*D^\mu\chi-(D^\mu\chi)^*\chi]$.

The static energy functional is given by
\begin{eqnarray}   \label{4.4}
E&=&\int\! d^2\!{\bf r}\, \left\{|D_0\phi|^2+|D_0\chi|^2
       +|D_i\phi|^2+|D_i\chi|^2+U(\phi,\chi)\right\}\nonumber\\
 &=&\int\! d^2\!{\bf r}\, \left\{|D_i\phi|^2+|D_i\chi|^2
  +\frac{1}{4|\phi|^2}
  \left[\frac{q_2^2\kappa_1b^1-q_2^1\kappa_2b^2}{q_1^1q_2^2-q_2^1q_1^2}
  \right]^2 \right.\nonumber\\
& &\hspace{13mm}\left.+\frac{1}{4|\chi|^2}\left[
     \frac{q_1^2\kappa_1b^1-q_1^1\kappa_2b^2}{q_1^1q_2^2-q_2^1q_1^2}\right]^2
    +U(\phi,\chi)\right\}\,,
\end{eqnarray}
where, on the second line, we have used the following relations (derived from
the Gauss laws (\ref{4.3}), assuming time-independent fields):
\begin{eqnarray}   \label{4.5}
q_1^1a_0^1+q_1^2a_0^2&=&
 -\frac{(q_2^2\kappa_1)b^1-(q_2^1\kappa_2)b^2}{2(q_1^1q_2^2-q_2^1q_1^2)
 |\phi|^2}\,,\nonumber\\
q_2^1a_0^1+q_2^2a_0^2
&=&-\frac{(q_1^2\kappa_1)b^1-(q_1^1\kappa_2)b^2}{2(q_1^1q_2^2-q_2^1q_1^2)
   |\chi|^2}\,.
\end{eqnarray}
Now suppose that the potential $U$ has the form
\begin{equation}  \label{4.6}
U=|\phi|^2\left\{\sigma_1\beta_{11}(|\phi|^2-c^2)
  +\sigma_2\beta_{12}(|\chi|^2-{c'}^2)\right\}^2
  +|\chi|^2\left\{\sigma_1\beta_{21}(|\phi|^2-c^2)
  +\sigma_2\beta_{22}(|\chi|^2-{c'}^2)\right\}^2,
\end{equation}
where $\sigma_p=+1$ or $-1$ ($p=1,2$), the $\beta_{pp'}$'s are defined as in
Eq. (\ref{2.15}) (i.e., $\beta_{11}=\frac{(q_1^1)^2}{\kappa_1}
+\frac{(q_1^2)^2}{\kappa_2}$,
$\beta_{12}=\frac{q_1^1q_2^1}{\kappa_1}+\frac{q_1^2q_2^2}{\kappa_2}$, etc.),
and $c$, $c'$ are arbitrary constants. Then, with the help of the identities
analogous to Eq. (\ref{3.6}), it is straightforward to show that the energy
functional in Eq. (\ref{4.4}) can be rewritten as
\begin{eqnarray}   \label{4.7}
E&=&\int\! d^2\!{\bf r}\,\left\{\frac{1}{4|\phi|^2}\left[\left(
    \frac{q_2^2\kappa_1b^1-q_2^1\kappa_2b^2}{q_1^1q_2^2-q_2^1q_1^2}\right)
    +2|\phi|^2\left\{\sigma_1\beta_{11}(|\phi|^2-c^2)
    +\sigma_2\beta_{12}(|\chi|^2-{c'}^2)\right\}\right]^2\right.\nonumber\\
 &&\hspace{14mm}+\frac{1}{4|\chi|^2}\left[\left(
    \frac{q_2^1\kappa_1b^1-q_1^1\kappa_2b^2}{q_1^1q_2^2-q_2^1q_1^2}\right)
    +2|\chi|^2 \left\{\sigma_1\beta_{21}(|\phi|^2-c^2)
    +\sigma_2\beta_{22}(|\chi|^2-{c'}^2)\right\}\right]^2\nonumber\\
 &&\hspace{14mm}+|D_1\phi+i\sigma_1D_2\phi|^2
   +|D_1\chi+i\sigma_2D_2\chi|^2\nonumber\\
 &&\hspace{14mm}\left.
   +\sigma_1c^2(q_1^1b^1+q_1^2b^2)+\sigma_2{c'}^2(q_2^1b^1+q_2^2b^2)\right\}\,.
\end{eqnarray}
Hence, for the theory defined by the Lagrangian density (\ref{4.1}) with the
potential (\ref{4.6}), there exists a Bogomol'nyi-type bound \cite{26} for the
static energy (which is nonnegative),
\begin{equation}
E\ge\sigma_1c^2\Phi_\phi+\sigma_2{c'}^2\Phi_\chi\,,
\end{equation}
where the fluxes $\Phi_\phi$ and $\Phi_\chi$ are defined as in Eq.
(\ref{3.19}).
Since $E$ must be nonnegative, this bound is meaningful only for a positive
value of $\sigma_1c^2\Phi_\phi+\sigma_2{c'}^2\Phi_\chi$.

Of particular interest here
are the solutions saturating the above Bogomol'nyi bound, which are realized
if and only if all squared expressions appearing in the integrand of Eq.
(\ref{4.7}) vanish identically. This gives rise to the following self-duality
equations:
\begin{equation}   \label{4.11}
D_1\phi+i\sigma_1D_2\phi=0\,,\hspace{5mm}D_1\chi+i\sigma_2D_2\chi=0\,,
\end{equation}
and
\begin{eqnarray}   \label{4.12}
b^1&=&-\frac{2}{\kappa_1}q_1^1|\phi|^2 [\sigma_1\beta_{11}
      (|\phi|^2-c^2)+\sigma_2\beta_{12}(|\chi|^2-{c'}^2)]\nonumber\\
   & &-\frac{2}{\kappa_2}q_2^1|\chi|^2 [\sigma_1\beta_{21}
      (|\phi|^2-c^2)+\sigma_2\beta_{22}(|\chi|^2-{c'}^2)]\,,\nonumber\\
b^2&=&-\frac{2}{\kappa_1}q_1^2|\phi|^2 [\sigma_1\beta_{11}
      (|\phi|^2-c^2)+\sigma_2\beta_{12}(|\chi|^2-{c'}^2)]\nonumber\\
   & &-\frac{2}{\kappa_2}q_2^2|\chi|^2 [\sigma_1\beta_{21}
      (|\phi|^2-c^2)+\sigma_2\beta_{22}(|\chi|^2-{c'}^2)]\,.
\end{eqnarray}
We expect that, for the parameters in some range at least, this system of
equations admit both topological and non-topological soliton solutions just as
in the model of Refs. \cite{22,23}. To analyze these equations, we may again
write $a_i^I({\bf r})$ as in Eq. (\ref{3.8}) and then, for the functions
$f_1({\bf r})$ and $f_2({\bf r})$ defined by
\begin{equation}   \label{4.13}
\phi({\bf r})=e^{-\sigma_1\{q_1^1U^1+q_1^2U^2\}}f_1({\bf r})\,,\hspace{4mm}
\chi({\bf r})=e^{-\sigma_2\{q_2^1U^1+q_2^2U^2\}}f_1({\bf r})\,,
\end{equation}
Eq. (\ref{4.11}) reduces to the statement of complex analyticity, i.e.,
$f_1=f_1(z_{(\sigma_1)})$ and $f_2=f_2(z_{(\sigma_2)})$
with $z_{(\sigma_p)}=x+i\sigma_py$. At the same time,
we use Eq. (\ref{4.13}) to express $b^1({\bf r})$ and $b^2({\bf r})$
in terms of $|\phi|$ and $|\chi|$, and then combine them with Eq. (\ref{4.12}).
The results are the following equations\footnote{Notice that the
$q$'s, $\kappa_1$ and $\kappa_2$ enter Eq. (\ref{4.14}) only through the
quantities $\beta_{pp'}$. This is an expected result even in the present
relativistic case, for the given equations
should remain the same under suitable linear transformations on the C-S
fields.}:
\begin{eqnarray}  \label{4.14}
\nabla^2\ln|\phi|^2
   &=&4\sigma_1\left\{\beta_{11}|\phi|^2 [\sigma_1\beta_{11}
      (|\phi|^2-c^2)+\sigma_2\beta_{12}(|\chi|^2-{c'}^2)]\right. \nonumber\\
   & &\hspace{8mm}\left.+\beta_{12}|\chi|^2 [\sigma_1\beta_{21}
      (|\phi|^2-c^2)+\sigma_2\beta_{22}(|\chi|^2-{c'}^2)]\right\}
       +4\pi\sum_{r=1}^{n_1}\delta^2({\bf r}-{\bf R}_r)\,,\nonumber\\
\nabla^2\ln|\chi|^2
   &=&4\sigma_2\left\{\beta_{21}|\phi|^2 [\sigma_1\beta_{11}
      (|\phi|^2-c^2)+\sigma_2\beta_{12}(|\chi|^2-{c'}^2)]\right.\nonumber\\
   & &\hspace{8mm}\left.+\beta_{22}|\chi|^2 [\sigma_1\beta_{21}
      (|\phi|^2-c^2)+\sigma_2\beta_{22}(|\chi|^2-{c'}^2)]\right\}
      +4\pi\sum_{r=1}^{n_2}\delta^2({\bf r}-{\bf R}'_r)\,.\nonumber\\
   & &
\end{eqnarray}
We have here assumed that the fields $\phi$ and $\chi$ have zeros at
$({\bf R}_1,\ldots,{\bf R}_{n_1})$ and $({\bf R}'_1,\ldots,{\bf R}'_{n_2})$,
respectively.
Notice that there is a certain similarity between Eq. (\ref{4.14}) and Eq.
(\ref{3.13}). This is not surprising since one can recover the model discussed
in Sec.III as the non-relativistic limit \cite{11} of this relativistic theory
(restricted to the nontopological soliton sector).

A general investigation on possible solutions to Eq. (\ref{4.14}) is beyond
the scope of this paper. We will below concentrate on a particularly
interesting special case, the self-dual system with $\beta_{11}=\beta_{22}=0$
and $\beta_{12}\neq0$. Note that we studied the non-relativistic model
under the same
condition in Eqs. (\ref{3.24})--(\ref{3.26}). Choosing the parameters as in
Eq. (\ref{3.24}), Eq. (\ref{4.14}) assumes a much simpler form, viz.,
\begin{eqnarray}  \label{4.16}
\nabla^2\ln|\phi|^2
 &=&\frac{4q^2{q'}^2}{\kappa^2}|\chi|^2(|\phi|^2-c^2)\,,\nonumber\\
\nabla^2\ln|\chi|^2
 &=&\frac{4q^2{q'}^2}{\kappa^2}|\phi|^2(|\chi|^2-{c'}^2)\,,
\end{eqnarray}
with the $\delta$-function terms not written out explicitly.
The Lagrangian density for this system
reads
\begin{eqnarray}
{\cal L}&=&\kappa\epsilon^{\mu\nu\lambda}v_\mu^{(1)}\partial_\nu
       v_\lambda^{(2)}-|(\partial_\mu-iqv_\mu^{(1)})\phi|^2
       -|(\partial_\mu-iq'v_\mu^{(2)})\chi|^2\nonumber\\
    & &-\frac{q^2{q'}^2}{\kappa^2}|\phi|^2(|\chi|^2-{c'}^2)^2
       -\frac{q^2{q'}^2}{\kappa^2}|\chi|^2(|\phi|^2-c^2)^2\,
\end{eqnarray}
with the Gauss laws (for time-independent fields) given by
\begin{eqnarray}   \label{4.18}
\epsilon^{ij}\nabla_iv_j^{(1)}&=&-\frac{q'}{\kappa}J_\chi^0
                        =\frac{2{q'}^2}{\kappa}v^{(2)0}|\chi|^2\,,\nonumber\\
\epsilon^{ij}\nabla_iv_j^{(2)}&=&-\frac{q'}{\kappa}J_\phi^0
                        =\frac{2q^2}{\kappa}v^{(1)0}|\phi|^2\,.
\end{eqnarray}
There are two distinct classes of solutions to the self-duality equations.
The first is a topological soliton with the asymptotic behavior
\begin{equation}
\rightarrow\infty:\hspace{4mm}
 \frac{|\phi({\bf r})|}{c}\rightarrow 1,\hspace{4mm}
                \frac{|\chi({\bf r})|}{c'}\rightarrow 1,
\end{equation}
and the fluxes are quantized for this solution, i.e.,
$\Phi_\phi=q\int\! d^2\!{\bf r}\,\epsilon^{ij}\nabla_iv_j^{(1)}=2\pi n_1$ and
$\Phi_\chi=q'\int\! d^2\!{\bf r}\,\epsilon^{ij}\nabla_iv_j^{(2)}=2\pi n_2$
($n_1$, $n_2$: integers). The second class is a nontopological
soliton with the asymptotic behavior
\begin{equation}
\rightarrow\infty:\hspace{4mm}
 |\phi({\bf r})|\rightarrow \frac{{\rm const}}{r^{\alpha_1}},\hspace{4mm}
                |\chi({\bf r})|\rightarrow \frac{{\rm const}}{r^{\alpha_2}},
\end{equation}
($\alpha_1$ and $\alpha_2$ are real numbers larger than 1). For this
nontopological soliton the fluxes are not quantized: we here have the formulas
$\Phi_\phi=2\pi(n_1+\alpha_1)$ and $\Phi_\chi=2\pi(n_2+\alpha_2)$,
when the field $\phi(\chi)$ has vorticity $n_1(n_2)$.  A topological
soliton with $\Phi_\phi=2\pi n_1$ and $\Phi_\chi=2\pi n_2$ may conveniently be
visualized as an assembly of $|n_1|$ `$\phi$-vortices' with
respective centers at the zeros of $\phi$ and $|n_2|$ `$\chi$-vortices'
with respective centers at the zeros of $\chi$.

Notice that, for the above soliton configurations, the charges
$Q_\phi\equiv\int\! d^2\!{\bf r}\,
J_\phi^0$ and $Q_\chi\equiv \int\! d^2\!{\bf r}\, J_\chi^0$ are simply
related to the fluxes as
\begin{equation}
\Phi_\phi=-\frac{qq'}{\kappa}Q_\chi\,,\hspace{1cm}
\Phi_\chi=-\frac{qq'}{\kappa}Q_\phi\,,
\end{equation}
due to the Gauss laws (\ref{4.18}). This relationship suggests the existence
of {\it mutual\/} statistical interaction between $\phi$-vortices and
$\chi$-vortices; but, an assembly of $\phi$-vortices (or $\chi$-vortices) only
will show no peculiar statistical effect. This conclusion is further supported
by calculating the angular momentum
$J\equiv\int\! d^2\!{\bf r}\, \epsilon^{ij}x_iT^{0j}$, where
$T^{0j}$ denotes the momentum density in the theory. In fact, at least for a
spherically symmetric solution based on the form
\begin{eqnarray}
\phi({\bf r})&=&f(r)e^{in_1\theta}\,,\hspace{20mm}
         \chi({\bf r})=g(r)e^{in_2\theta}\,,\nonumber\\
qv_i^{(1)}({\bf r})&=&\epsilon^{ij}\frac{x^j}{r^2}[h^{(1)}(r)-n_1]\,,
\hspace{5mm}q'v_i^{(2)}({\bf r})=\epsilon^{ij}\frac{x^j}{r^2}[h^{(2)}(r)-n_2]
\end{eqnarray}
with
\begin{eqnarray}
&&h^{(1)}(\infty)=h^{(2)}(\infty)=0\,,\hspace{15mm}
                            \mbox{(topological soliton)}\nonumber\\
&&h^{(1)}=\alpha_2\,,\ \ \ h^{(2)}(\infty)=\alpha_1\,,\hspace{6mm}
                            \mbox{(nontopological soliton)}
\end{eqnarray}
a simple calculation gives the result
\begin{equation}
J=\left\{\begin{array}{ll}{\displaystyle
       -\frac{\pi\kappa}{qq'}n_1n_2\,,}&\mbox{(topological)}\\
      {\displaystyle -\frac{\pi\kappa}{qq'}(n_1n_2-\alpha_1\alpha_2)\,,}\ \ \ &
            \mbox{(nontopological)}\,. \end{array} \right.
\end{equation}
Thus, individual $\phi$- or $\chi$-vortices do not carry angular momentum. On
the other hand, a composite of a $\phi$-vortex and a $\chi$-vortex each has a
nonvanishing $J$, and this is an anticipated result in a system with mutual
statistical interaction.

The general solution to the given self-duality equations is difficult to
obtain, although certain subclasses of solutions can be readily identified in
terms of those of the previously known self-dual system. By evaluating the
index of the differential operator associated with the appropriate fluctuation
equation, the number of free parameters entering
the general solution with given
values of $\Phi_\phi$ and $\Phi_\chi$ is determined as
\begin{equation}  \label{4.25}
N=\left\{\begin{array}{ll}2n_1+2n_2\,,&\mbox{(topological)}\\
    2n_1+2\hat{\alpha}_1+2n_2+2\hat\alpha_2\,,\ \ \ &\mbox{(nontopological)}\,.
         \end{array} \right.
\end{equation}
Here the general topological soliton solution with $n_2=0$ but $n_1\neq0$ (or,
if one wishes, with $n_1=0$ but $n_2\neq0$) is easy to describe---one may set
$\chi({\bf r})=c'$ and, in view of Eq. (\ref{4.16}), just
choose $\phi({\bf r})$ to be a
solution to the familiar equation from the study of the Ginzburg-Landau-type
model \cite{26},
\begin{equation}
\nabla^2\ln|\phi|^2=\frac{4q^2{q'}^2{c'}^2}{\kappa^2}(|\phi|^2-c^2)\,.
\end{equation}
In this case, it follows from Eq. (\ref{4.18}) that
$v^{(1)0}({\bf r})=v_j^{(2)}({\bf r})=0$, while
$v^{(2)0}=\frac{\kappa}{2{q'}^2c'}\epsilon^{ij}
 \nabla_iv_j^{(1)}({\bf r})\neq0$. Another
subclass of topological or nontopological soliton solutions are obtained by
setting $|\chi({\bf r})|^2=\left(\frac{c'}{c}\right)^2|\phi({\bf r})|^2$,
and this of course corresponds to the case with $n_1=n_2$
(and $\alpha_1=\alpha_2$). For the
latter, the two equations in Eq. (\ref{4.16}) collapse to one, namely,
\begin{equation}  \label{4.27}
\nabla^2\ln|\phi|^2
=\frac{4q^2{q'}^2{c'}^2}{\kappa^2c^2}|\phi|^2(|\phi|^2-c^2)\,,
\end{equation}
the form of which matches precisely the corresponding equation encountered in
the study of the `minimal' self-dual C-S Higgs model \cite{22}. But
this does not comprise the full general solution in the sector specified by
$n_1=n_2$ (and $\alpha_1=\alpha_2$). The number of free parameters which
enter the
solution based on Eq. (\ref{4.27}) (as calculated in Ref. \cite{23}) is just a
half of the value given in Eq. (\ref{4.25}). This may be understood by
observing that the basic unit in a solution satisfying the condition
$|\chi|^2=\left(\frac{c'}{c}\right)^2|\phi|^2$ is assumed by ``a
$\phi$-vortex {\it on
top of\/} a $\chi$-vortex'', while the index theorem suggests the existence of
more general solution in which $\phi$-vortices and $\chi$-vortices serve as
separate units.

\section{SUMMARY AND DISCUSSION}
The precise nature of the Schr\"{o}dinger quantum field theory with general
$[U(1)]^N$ C-S interations has been clarified, a novel feature being the
existence of mutual statistical interactions between distinguishable
particles. Then, for the corresponding self-dual models with two matter
fields, we investigated the structure of classical soliton-type solutions to
the static equations of motion, with or without uniform external magnetic
field. In particular, to obtain a system which admits non-trivial
soliton solutions satisfying the self-duality equations, we
derived a set of necessary conditions for the parameters of the theory. While
our self-duality equations reduce to the Toda-type equations or their
generalizations, the matrix $K$ in the equations is not necessarily equal to
the Cartan matrix of a certain Lie algebra. For some special cases we
exhibited soliton solutions in a more explicit way. Soliton solutions in a
relativistic self-dual system with two C-S gauge fields were also discussed
briefly. We conjectured that these solitons exhibit mutual (as well as
fractional) statistics.

Some comments are in order. First of all, it is intriguing that the Toda
equation retains some of its interesting mathematical properties (e.g. the
existence of multi-soliton solutions) even if its structure is suitably
altered. Aside from that the matrix $K$ in Eq. (\ref{1.2}) need not have a
group theoretical origin, we saw this phenomenon realized when we add
constant terms on the right hand side (as in Eq. (\ref{3.38})) and also
quadratic terms in the densities (as in Eq. (\ref{4.12})). Quite possibly,
certain universal mathematical structures might exist behind all these models.
Also desirable will be to clarify further various physical properties (e.g.
statistics) of the vortex solitons discussed in this paper and to study their
possible roles in the {\it real\/} physics of multi-layered Hall media.
Another fruitful line of research is the quantum mechanical investigation of
our model Hamiltonian in Eq. (\ref{3.2}). We noted already that, by exploiting
the supersymmetry in this system, it should be possible to find the
corresponding {\it exact\/} ground state and also their degeneracy. Just as in
the case of a one layer system \cite{9}, this investigation might yield some
valuable insights in understanding the multi-layered fractional quantum Hall
effects.

\acknowledgments
This work (C.K., C.L., B.-H.L, and H.M.) was supported in part by the Korea
Science and Engineering Foundation (through the Center for Theoretical
Physics, Seoul National University) and also by the Ministry of Education,
Korea. The work of P. K. was supported by Department of Energy grant
\#DOE-DE-AC02-83ER-40105. B.-H.L. wishes to thank the Theoretical
Physics Institute of University of Minnesota for the hospitality
during his stay. He also thanks Y. Hosotani for helpful discussions.

\appendix
\section{}

Here we will first explain how Eqs. (\ref{2.27}) and (\ref{2.30}) are derived,
and then go on to establish the expression (\ref{2.31}). For the contribution
$B$ defined by Eq. (\ref{2.21}), one may repeatedly use the relations like
\begin{eqnarray}
\lefteqn{\Psi_p({\bf r}_{k-1}^{(p)},t)
  \left(-\frac{\hbar^2}{2m_p}\left[\nabla_i^{(p,k)}-\frac{i}{\hbar c}
  \epsilon^{ij}\nabla_j^{(p,k)}\sum_{p'}\beta_{pp'}
  \int\! d^2\!{\bf r}'\, G({\bf r}_k^{(p)}-{\bf r}')\rho_{p'}
  ({\bf r}',t)\right.\right.}\nonumber\\
& &\left.\left.\hspace{20mm}-\frac{i}{\hbar c}e_pA^{\rm ex}_i
  ({\bf r}_k^{(p)},t) \right]^2\Psi_p({\bf r}_k^{(p)},t)\right)\nonumber\\
&=&-\frac{\hbar^2}{2m_p}\left[\nabla_i^{(p,k)}
  -\frac{i}{\hbar c}\epsilon^{ij}\nabla_j^{(p,k)}
  \sum_{p'}\beta_{pp'}\int\! d^2\!{\bf r}'\,
  G({\bf r}_k^{(p)}-{\bf r}')\rho_{p'}\right.\nonumber\\
&&\hspace{10mm}\left.-\frac{i}{\hbar c}\epsilon^{ij}
  \nabla_j^{(p,k)}\beta_{pp} G({\bf r}_k^{(p)}-{\bf r}_{k-1}^{(p)})
  -\frac{i}{\hbar c}e_pA^{\rm ex}_i({\bf r}_k^{(p)},t)\right]^2
  \Psi_p({\bf r}_{k-1}^{(p)},t)\Psi_p({\bf r}_k^{(p)},t),\nonumber\\
&&
\end{eqnarray}
which follows from the noncommutativity of $\rho_{p'}({\bf r}',t)$ and
$\Psi_p({\bf r}_{k-1}^{(p)},t)$. Once all the field operators on the left
of squared
differential operator are relocated to its right by this procedure, one
readily recognizes the expression in Eq. (\ref{2.27}) as a consequence of the
definition (\ref{2.18}) and the fact that $\langle0|\rho_{p'}({\bf r}',t)=0$.
For the
contribution $A$, more steps are necessary to derive the result (\ref{2.30})
from the form in Eq. (\ref{2.20}). Here, using the commutation relations
(\ref{2.2}) and (\ref{2.13}), we first observe that
\begin{eqnarray} \label{2.28}
\lefteqn{\Psi_p({\bf r}_{k-1}^{(p)},t)\left(\sum_{p'}\frac{\beta_{pp'}}{c}
    \epsilon^{ij}\nabla_j^{(p,k)}\int\! d^2\!{\bf r}'\,
    G({\bf r}_k^{(p)}-{\bf r}')J_{p'i}({\bf r}',t)\right)}\nonumber\\
 &=&\left(-\frac{i\hbar}{m_pc}\beta_{pp}
    \epsilon^{ij}\nabla_j^{(p,k)}G({\bf r}_k^{(p)}-{\bf r}_{k-1}^{(p)})
    D_i^{(p,k-1)}\right.\nonumber\\
 &&-\sum_{p'}\frac{1}{m_{p'}c^2}\beta_{pp'}^2\epsilon^{ij}\nabla_j^{(p,k)}
   \int\! d^2\!{\bf r}'\, G({\bf r}_k^{(p)}-{\bf r}')
   \epsilon^{il}\nabla_l^{({\bf r}')}G({\bf r}_{k-1}^{(p)}-{\bf r}')
   \Psi_{p'}^\dagger({\bf r}',t)\Psi_{p'}({\bf r}',t)\nonumber\\
 &&\left.+\sum_{p'}\frac{\beta_{pp'}}{c}\epsilon^{ij}\nabla_j^{(p,k)}
   \int\! d^2\!{\bf r}'\, G({\bf r}_k^{(p)}-{\bf r}')
   J_{p'i}({\bf r}',t)\right)\Psi_p({\bf r}_{k-1}^{(p)},t),
\end{eqnarray}
where $D_i^{(p,k)}$ is defined as in the case of $\nabla_i^{(p,k)}$. If we
further let the operator $\Psi_p({\bf r}_{k-2}^{(p)},t)$
act on the expression (\ref{2.28})
from the left, the result can then be written as
\begin{eqnarray}
&&\hspace{-2mm}\left(-\frac{i\hbar}{m_p c}\beta_{pp}
  \epsilon^{ij}\nabla_j^{(p,k)} G({\bf r}_k^{(p)}-{\bf r}_{k-1}^{(p)})
  D_i^{(p,k-1)}-\frac{i\hbar}{m_p c}\beta_{pp} \epsilon^{ij}\nabla_j^{(p,k)}
  G({\bf r}_{k}^{(p)}-{\bf r}_{k-2}^{(p)})D_i^{(p,k-2)}\right.\nonumber\\
&&-\frac{i\hbar}{m_p c^2}\beta_{pp}^2\epsilon^{ij}\nabla_j^{(p,k)}
   G({\bf r}_k^{(p)}-{\bf r}_{k-1}^{(p)}) \epsilon^{il}\nabla_l^{(p,k-1)}
   G({\bf r}_{k-1}^{(p)}-{\bf r}_{k-2}^{(p)})\nonumber\\
&&-\frac{i\hbar}{m_p c^2}\beta_{pp}^2\epsilon^{ij}\nabla_j^{(p,k)}
   G({\bf r}_k^{(p)}-{\bf r}_{k-2}^{(p)}) \epsilon^{il}\nabla_l^{(p,k-2)}
   G({\bf r}_{k-2}^{(p)}-{\bf r}_{k-1}^{(p)})\nonumber\\
&&-\sum_{p'}\frac{1}{m_{p'}c^2}\beta_{pp'}^2\epsilon^{ij}\nabla_j^{(p,k)}
   \int\! d^2\!{\bf r}'\, G({\bf r}_k^{(p)}-{\bf r}')\epsilon^{il}
   \nabla_l^{({\bf r}')}[G({\bf r}_{k-1}^{(p)}-{\bf r}')
   +G({\bf r}_{k-2}^{(p)}-{\bf r}')]\nonumber\\
&&\hspace{50mm}\Psi_{p'}^\dagger({\bf r}',t)\Psi_{p'}({\bf r}',t)\nonumber\\
&&\left.+\sum_{p'}\frac{\beta_{pp'}}{c}\epsilon^{ij}\nabla_j^{(p,k)}
   \int\! d^2\!{\bf r}'\, G({\bf r}_k^{(p)}-{\bf r}')J_{p'i}
   ({\bf r}',t)\right)\Psi_p({\bf r}_{k-2},t)\Psi_p({\bf r}_{k-1}^{(p)},t),
\end{eqnarray}
where we have again used Eq. (\ref{2.13}). Now it is not difficult to infer
that as analogous steps are repeatedly all the way, the final result should be
the expression (\ref{2.30}).

To show that the sum of $A$, $B$ and $D$ can be expressed as in Eq.
(\ref{2.31}), we proceed as follows. We begin with the trivial observation,
$\displaystyle\sum_{(p',k')\neq(p,k)}
  =\sum_{(p',k')<(p,k)}+\sum_{(p',k')>(p,k)}$, to cast the right hand side of
Eq. (\ref{2.31}) into the form
\begin{eqnarray}\label{2.32}
B&+&\sum_{(p,k)}\frac{i\hbar}{m_p c}\epsilon^{ij}\nabla_j^{(p,k)}
\left(\sum_{(p',k')>(p,k)}\beta_{pp'}G({\bf r}_k^{(p)}-{\bf r}_{k'}^{(p')})
 \right)\nonumber\\
 &&\hspace{4mm}\cdot\left[\nabla_i^{(p,k)}-\frac{i}{\hbar c}
  \epsilon^{il}\nabla_l^{(p,k)} \left(\sum_{(p'',k'')<(p,k)}
  \beta_{pp''}G({\bf r}_k^{(p)}-{\bf r}_{k''}^{(p'')})\right)
  -\frac{i}{\hbar c}e_pA^{\rm ex}_i({\bf r}_k^{(p)},t)\right]\nonumber\\
 &&\hspace{4mm}\cdot\Phi({\bf r}_1^{(1)},
   \ldots,{\bf r}_{n_M}^{(M)},t)\nonumber\\
 &+&\sum_{(p,k)}\sum_{(p',k')>(p,k)}\sum_{(p'',k'')>(p,k)}
    \frac{1}{2m_pc^2}\beta_{pp'}\beta_{pp''}\left(\epsilon^{ij}\nabla_j^{(p,k)}
    G({\bf r}_k^{(p)}-{\bf r}_{k'}^{(p')})\right)\nonumber\\
 &&\hspace{40mm}\cdot\left(\epsilon^{il}\nabla_l^{(p,k)}G({\bf r}_k^{(p)}
    -{\bf r}_{k''}^{(p'')})\right)\Phi({\bf r}_1^{(1)},
    \ldots,{\bf r}_{n_M}^{(M)},t).
\end{eqnarray}
The second term in this expression can then be rewritten as
\begin{eqnarray}\label{2.33}
\lefteqn{-\sum_{(p,k)}\sum_{(p',k')<(p,k)}\frac{i\hbar}{m_{p'}c}
         \epsilon^{ij}\nabla_j^{(p,k)}\beta_{pp'}
         G({\bf r}_{k'}^{(p')}-{\bf r}_k^{(p)})}\nonumber\\
&&\hspace{13mm}\cdot\left[\nabla_i^{(p',k')}
 -\frac{i}{\hbar c}\epsilon^{ij}\nabla_j^{(p',k')}
 \left(\sum_{(p'',k'')<(p',k')}\beta_{p'p''}
 G({\bf r}_k^{(p')}-{\bf r}_{k''}^{(p'')}) \right)
 -\frac{i}{\hbar c}e_pA^{\rm ex}_i({\bf r}_{k'}^{(p')},t)\right]\nonumber\\
 &&\hspace{13mm}\cdot\Phi({\bf r}_1^{(1)},\ldots,{\bf r}_{n_M}^{(M)},t)
\end{eqnarray}
while the last term is equal to
\begin{eqnarray}  \label{2.34}
&&\sum_{(p,k)}\sum_{(p',k')<(p,k)}\frac{1}{2m_{p'}c^2}\beta_{p'p}^2\left[
 \epsilon^{ij}\nabla_j^{(p',k')}G({\bf r}_k^{(p)}-{\bf r}_{k'}^{(p')})\right]^2
 \Phi({\bf r}_1^{(1)},\ldots,{\bf r}_{n_M}^{(M)},t) \nonumber\\
&&\hspace{3mm}-\sum_{(p,k)}\sum_{(p',k')<(p,k)}
              \sum_{(p'',k'')>(p',k')\atop (p'',k'')>(p,k)}
\frac{1}{m_{p'}c^2}\beta_{p'p}\beta_{p'p''}\left(\epsilon^{ij}\nabla_j^{(p,k)}
  G({\bf r}_k^{(p)}-{\bf r}_{k'}^{(p')})\right)\nonumber\\
&&\hspace{40mm}\cdot\left(\epsilon^{il}\nabla_l^{(p,k)}
  G({\bf r}_k^{(p)}-{\bf r}_{k''}^{(p'')})
  \right)\Phi({\bf r}_1^{(1)},\ldots,{\bf r}_{n_M}^{(M)},t).
\end{eqnarray}
The first term in Eq. (\ref{2.34}) evidently coincides with the contribution
$D$ shown in Eq. (\ref{2.25}). On the other hand, the second term in
Eq. (\ref{2.34}) and the expression
(\ref{2.33}) combine to yield the contribution $A$ in Eq. (\ref{2.30}).
The equation (\ref{2.31}) has been established now.

\section{}

As regards the solutions to the self-duality equations (\ref{3.7a}) and
(\ref{3.7b}), we will here present the index theorem analysis. Our immediate
concern is to count the free parameters entering the general soliton solution,
with given flux values $\sum_Iq_p^I\int\! d^2\!{\bf r}\, b^I({\bf r})
 =\sigma_p2\pi\hbar c(n_p+\alpha_p)$
$(p=1,\ldots,M)$ and the asymptotic behaviors
\begin{equation}
r\longrightarrow\infty :\hspace{4mm}
|\Psi_p({\bf r})|\longrightarrow\frac{{\rm const}}{r^{\alpha_p}}\,,
\hspace{4mm} (\alpha_p>1)\,.
\end{equation}
Thanks to Eq. (\ref{3.11}), the integer $n_p$ here coincides with the number
of zeros for the field $\Psi_p({\bf r})$. The number of free parameters in
question
is equal to the number of normalizable zero modes of the small fluctuation
equations given in the background of any specific soliton solution. From Eqs.
(\ref{3.7a}) and (\ref{3.7b}), the fluctuation equations are of the following
form:
\begin{mathletters}
\begin{eqnarray}  \label{B.2a}
&&(D_1+i\sigma_pD_2)\delta\Psi_p
 -\frac{i}{\hbar c}\Psi_p\sum_Iq_p^I(\delta a_1^I+i\sigma_p\delta a_2^I)=0\,,\\
&&\sum_Iq_p^I(\nabla_1\delta a_2^I
     -\nabla_2\delta a_1^I)\sum_{P'}\beta_{pp'}
    (\Psi_p^*\delta\Psi_p+\Psi_p\delta\Psi_p^*)=0\,.
      \label{B.2b}
\end{eqnarray}
\end{mathletters}
[Here, $(q_p^I)$ is being supposed to be a non-singular $M\times M$ matrix;
but note that our formula (\ref{B.9}) is valid for more general $(q_p^I)$.]
Then, to eliminate superficial zero modes related to the freedom of local
gauge transformations, we may generalize the real equation (\ref{B.2b}) to the
complex equation
\begin{equation}  \label{B.3}
(\nabla_1-i\sigma_p\nabla_2)\sum_Iq_p^I(\sigma_p a_1^I+i\sigma_p\delta a_2^I)
  +2i\sigma_p\sum_{p'}\beta_{pp'}\Psi_{p'}^*\delta\Psi_{p'}=0\,.
\end{equation}
Taking the imaginary part of this equation reproduces Eq. (\ref{B.2b}), while
the real part can be viewed as the gauge condition.

Equations (\ref{B.2a}) and (\ref{B.3}) are represented by a single matrix
equation
\begin{equation}   \label{B.4}
{\cal D}\left(\begin{array}{c}
    \delta\Psi_1\\ \vdots\\ \delta\Psi_M\\
    {[\sum_Iq_1^I\delta a_1^I]+i\sigma_1[\sum_Iq_1^I\delta a_2^I]}\\ \vdots\\
    {[\sum_Iq_M^I\delta a_1^I]+i\sigma_M[\sum_Iq_M^I\delta a_2^I]}
    \end{array}\right)=0\,,
\end{equation}
where
\begin{equation}
{\cal D}=\left(\begin{tabular}{ccc|ccc}
$D_1+i\sigma_1D_2$& &0&$-\frac{i}{\hbar c}\Psi_1$&  &0\\
  &$\ddots$& & &$\ddots$& \\
0& &$D_1+i\sigma_MD_2$&0& &$-\frac{i}{\hbar c}\Psi_M$\\ \hline
$2i\sigma_1\bar\beta_{11}\Psi_1^*$&$\ldots$&$2i\sigma_1
 \bar\beta_{1M}\Psi_M^*$&$\nabla_1-i\sigma_1\nabla_2$& &0\\
$\vdots$& &$\vdots$& &$\ddots$& \\
$2i\sigma_M\bar\beta_{M1}\Psi_1^*$&$\ldots$&$2i\sigma_M
 \bar\beta_{MM}\Psi_M^*$& 0 & &$\nabla_1-i\sigma_M\nabla_2$
\end{tabular} \right)
\end{equation}
The index of $\cal D$ is defined as
\begin{equation} \label{B.6}
\mbox{Index}\,({\cal D})=\dim (\mbox{kernel}\;{\cal D})
      -\dim (\mbox{kernel}\;{\cal D}^\dagger)\,.
\end{equation}
To calculate this index, it is convenient to consider the quantity \cite{27}
\begin{equation}
I(M^2)=\mbox{Tr}\left[\frac{M^2}{{\cal D}^\dagger{\cal D}+M^2}\right]
       -\mbox{Tr}\left[\frac{M^2}{{\cal D}{\cal D}^\dagger+M^2}\right]\,,
\end{equation}
which can be shown to be independent of $M^2$. Naively, one may expect to
recover the above index in the $M^2\rightarrow0$ limit. On the other hand, a
straightforward evaluation of $I(M^2)$ in the limit
$M^2\rightarrow\infty$ gives
\begin{eqnarray} \label{B.8}
I(M^2)&=&\frac{1}{2\pi\hbar c}\sum_{p,I}\sigma_p\int\! d^2\!{\bf r}\,
         q_p^Ib^I\nonumber\\
      &=&\sum_p(n_p+\alpha_p)\,.
\end{eqnarray}
Note that this is not integer-valued in general, while the index defined by
Eq. (\ref{B.6}) is necessarily an integer. This discrepancy is due to the
continuum spectrum extending to zero, which gives rise to a non-zero
contribution to $I(M^2)$. Subtracting this contribution from
the value in Eq. (\ref{B.8}) (see Refs. \cite{18,28}),
the correct value for the index reads
\begin{equation}  \label{B.9}
\mbox{Index}\,({\cal D})=\sum_p(n_p+\hat\alpha_p)\,,
\end{equation}
where $\hat\alpha_p$ denotes the largest integer less than $\alpha$.
Further, by the
manipulations analogous to those described in Refs. \cite{18,23}, it is not
difficult to show that $\dim(\mbox{kernel}\;{\cal D}^\dagger)=0$. So the kernel
of $\cal D$ has the (complex) dimension $\sum_{p}(n_p+\hat\alpha_p)$. Based on
this, we now conclude that the total number of (real) free parameters in the
general solution of the given character is equal to
$2\sum_p(n_p+\hat\alpha_p)$.
Also, in the case of topological soliton solutions which are allowed
in the presence of non-zero uniform externel magnetic field, it
suffices to delete the $\hat{\alpha}_p$-term in this result.

\begin{figure}
\caption{A closed curve $C$ formed as the positions of two identical
particles, ${\bf r}_{k_1}^{(p)}$ and ${\bf r}_{k_2}^{(p)}$, are exchanged.}
\end{figure}
\begin{figure}
\caption{For parameters lying in the shaded regions, there exists no
solution to the self-duality equations.}
\end{figure}
\begin{figure}
\caption{Non-Liouville-type solutions in the case of $\beta_{11}=\beta_{22}=0$.
(a)The plot of a solution with vorticities $n_1=n_2=0$. We have here chosen
$\phi(0)=0.55$ and $\chi(0)=0.5$.
(b)The plot of a solution with vorticities $n_1=0$ and $n_2=1$.}
\end{figure}

\begin{references}
\bibitem{1}J. M. Leinaas and J. Myrheim, Nuovo Cimento {\bf 37B}, 1 (1977);
   G. Goldin, R. Menikoff, and D. Sharp, J. Math. Phys. {\bf 21}, 650 (1980);
   F. Wilczek, Phys. Rev. Lett. {\bf 48}, 1144 (1982); {\bf 49}, 957 (1982).
\bibitem{2}R. Jackiw and S. Templeton, Phys. Rev. {\bf D23}, 2291 (1981);
 J. Schonfeld, Nucl. Phys. {\bf B185}, 157 (1981);
S. Deser, R. Jackiw and S. Templeton, Ann. Phys. (N. Y.) {\bf 140}, 372 (1982);
 D. Arovas, J. Schrieffer, F. Wilczek and A. Zee, Nucl. Phys. {\bf B251}, 117
   (1985);
   C. R. Hagen, Ann. Phys. (N. Y.) {\bf 157}, 342, (1984).
\bibitem{3}S. M. Girvin and A. H. MacDonald, Phys. Rev. Lett. {\bf 58},
 1252 (1987);
S. C. Zhang, T. Hansson, and S. Kivelson, Phys. Rev. Lett. {\bf 62}, 82 (1989);
   N. Read, Phys. Rep. {\bf 62}, 85 (1989).
\bibitem{4}X. G. Wen and A. Zee, Nucl. Phys. {\bf B15}, 135 (1990);
  Phys. Rev. {\bf B44}, 274(1991); Phys. Rev. {\bf B46}, 2290 (1992);
  B. Blok and X. G. Wen, Phys. Rev. {\bf B43}, 8337 (1991);
 J. Fr\"{o}hlich and A. Zee, Nucl. Phys. {\bf B364}, 517 (1991); D. Wesolowski,
  Y. Hosotani and C.-L. Ho, preprint UMN-TH-1128/93.
\bibitem{5}F. Wilczek, Phys. Rev. Lett. {\bf 69}, 132 (1992);
   C. R. Hagen, Phys. Rev. Lett. {\bf 68}, 3821 (1992).
\bibitem{6}Z. F. Ezawa and A. Iwazaki, Int. J. Mod. Phys. {\bf B6}, 3205
(1992).
\bibitem{7}C. Ting, Int. J. Mod. Phys. {\bf B6}, 3155 (1992).
\bibitem{8}Y. Aharonov and A. Casher, Phys. Rev. {\bf A19}, 2461 (1979).
\bibitem{9}S. M. Girvin, A. H, MacDonald, M. Fisher, S. J. Rey, and J. Sethna,
   Phys. Rev. Lett. {\bf 65}, 1671 (1990);
   M. Y. Choi, C. Lee, and J. Lee, Phys. Rev. {\bf B46}, 1489 (1992).
\bibitem{10}R. Jackiw and S.-Y. Pi, Phys. Rev. Lett. {\bf 64}, 2969 (1990).
\bibitem{11}R. Jackiw and S.-Y. Pi, Phys. Rev. {\bf D42}, 3500 (1990).
\bibitem{12}G. Dunne, R. Jackiw, S.-Y. Pi and C. Trugenberger,
  Phys. Rev. {\bf D43}, 1332 (1991).
\bibitem{13}G. Dunne, MIT preprint CTP \#2097, March 1992.
\bibitem{ho}C.-L. Ho and Y. Hosotani, Int. J. Mod. Phys. {\bf A7},
	    5797 (1992).
\bibitem{fetter}See, for instance, A. L. Fetter and J. D. Walecka,
     {\it Quantum Theory of Many-Particle Systems} (McGraw-Hill, New
     York, 1971).
\bibitem{14}Y. Aharonov and A. Casher, Phys. Rev. Lett. {\bf 53}, 319 (1984).
\bibitem{15}Z. F. Ezawa, M. Hotta and A. Iwazaki,
     Phys. Rev. {\bf D44}, 452 (1991);
    G. Dunne and C. Trugenberger, private communication;
    S. L. Sondhi and S. A. Kivelson, Phys. Rev. {\bf B46}, 13319 (1992).
\bibitem{16}M. Toda, Prog. Theor. Phys. (Suppl) {\bf 45}, 174 (1970);
    Phys. Rep. {\bf 18}, 1 (1975).
\bibitem{17}A. Leznov and M. Savaliev, Lett. Math. Phys. {\bf 3}, 489 (1979);
    Comm. Math. Phys. {\bf 74}, 111 (1980);
    B. Konstant, Avd. Math. {\bf 34}, 195 (1979).
\bibitem{18}S. K. Kim, K. S. Soh and J. H. Yee,
   Phys. Rev. {\bf D42}, 4139 (1990).
\bibitem{19}S. K. Kim, W. Namgung, K. S. Soh and J. H. Yee,
    Phys. Rev. {\bf D46}, 1882 (1992).
\bibitem{20}S. K. Kim and H.-S. Min, Phys. Lett. {\bf 281B}, 81 (1992);
    Y. Kim and K.  Lee, preprint(CU-TP-574), 1992.
\bibitem{21}C. Kim, Phys. Rev. {\bf D47}, 673 (1993).
\bibitem{22}J. Hong, Y. Kim and P.-Y. Pac, Phys. Rev. Lett.
   {\bf 64}, 2230 (1990);
   R. Jackiw and E. Weinberg, Phys. Rev. Lett. {\bf 64}, 2234 (1990).
\bibitem{23}R. Jackiw, K. Lee and E. Weinberg,
   Phys. Rev. {\bf D42}, 3488 (1990).
\bibitem{24}R. Wang, Comm. Math. Phys. {\bf 137}, 587 (1991).;
   J. Spruck and Yisong Yang, Comm. Math. Phys. {\bf 149}, 361 (1992).
\bibitem{25}T. Vachaspati and A. Ach\`{u}carro, Phys. Rev. {\bf D44},
 3067 (1991); M. Hindmarsh, Phys. Rev. Lett. {\bf 68}, 1263 (1992);
 G. W. Gibbons, M. E. Ortiz,
            F. Ruiz and T. M. Samols, Nucl. Phys. {\bf B385}, 127 (1992).
\bibitem{26}E. B. Bogomol'nyi, Yad. Fiz. {\bf 24}, 861 (1976) [Sov. J. Nucl.
           Phys. {\bf 24}, 449 (1976)].
\bibitem{27}L. S. Brown, R. D. Carlitz and C. Lee, Phys. Rev.
   {\bf D16}, 417 (1977); E. J.
   Weinberg, Phys. Rev. {\bf D19}, 3008 (1979); see also {\bf 24}, 2669 (1981).
\bibitem{28}K. Kiskis, Phys. Rev. {\bf D15}, 2329 (1977);
   M. Ansourian, Phys. Lett. {\bf 70B}, 301 (1977);
   D. Boyanovsky and R. Blankenbecler, Phys. Rev. {\bf D31}, 3234 (1985).
\end{references}
\end{document}